\documentclass[twocolumn,showpacs,preprintnumbers,superscriptaddress,amsmath,amssymb,aps,epsfig,nofootinbib]{revtex4}

\usepackage{graphicx}
\usepackage{dcolumn}
\usepackage{bm}
\usepackage{latexsym}
\usepackage{longtable}
\usepackage{enumerate}
\usepackage{subfloat}
\usepackage{array}
\usepackage{xcolor}
\usepackage{footmisc}

\renewcommand *\arraystretch{1.5}

\let\oldhat\hat
\renewcommand{\hat}[1]{\oldhat{\mathbf{#1}}}
\let\oldvec\vec
\renewcommand{\vec}[1]{\oldvec{\mathbf{#1}}}

\newcommand{\AltTable}{\setlength\LTcapwidth{3.5in}}

\begin{document}


\title{Radiative-capture cross sections for the $^{139}$La(n,$\gamma$) reaction using thermal neutrons and structural properties of $^{140}$La}

\author{A.~M.~Hurst} 
\email{amhurst@berkeley.edu}
\affiliation{Department of Nuclear Engineering, University of California, Berkeley, California 94720, USA}

\author{A.~Sweet}
\affiliation{Department of Nuclear Engineering, University of California, Berkeley, California 94720, USA}
\author{B.~L.~Goldblum}
\affiliation{Department of Nuclear Engineering, University of California, Berkeley, California 94720, USA}
\author{R.~B.~Firestone}
\affiliation{Department of Nuclear Engineering, University of California, Berkeley, California 94720, USA}
\author{M.~S.~Basunia}
\affiliation{Lawrence Berkeley National Laboratory, Berkeley, California 94720, USA}
\author{L.~A.~Bernstein}
\affiliation{Department of Nuclear Engineering, University of California, Berkeley, California 94720, USA}
\affiliation{Lawrence Berkeley National Laboratory, Berkeley, California 94720, USA}
\author{Zs.~R{\'e}vay} 
\affiliation{Centre for Energy Research, Hungarian Academy of Sciences, H-1525 Budapest, Hungary}
\affiliation{Technische Universit{\"a}t M{\"u}nchen, Forschungsneutronenquelle Heinz Maier-Leibnitz (FRM II), Garching, Germany}
\author{L.~Szentmikl{\'o}si} 
\affiliation{Centre for Energy Research, Hungarian Academy of Sciences, H-1525 Budapest, Hungary}
\author{T.~Belgya}
\affiliation{Centre for Energy Research, Hungarian Academy of Sciences, H-1525 Budapest, Hungary}
\author{J.~E.~Escher}
\affiliation{Lawrence Livermore National Laboratory, Livermore, California 94550, USA}
\author{I.~Hars{\'a}nyi}
\affiliation{Centre for Energy Research, Hungarian Academy of Sciences, H-1525 Budapest, Hungary}
\author{M.~Krti{\v c}ka}
\affiliation{Charles University in Prague, Faculty of Mathematics and Physics, CZ-180 00 Prague, Czech Republic}
\author{B.~W.~Sleaford}
\affiliation{Lawrence Livermore National Laboratory, Livermore, California 94550, USA}
\author{J.~Vujic}
\affiliation{Department of Nuclear Engineering, University of California, Berkeley, California 94720, USA}

\date{\today}

\begin{abstract}
A set of prompt partial $\gamma$-ray production cross sections from thermal neutron-capture were measured for the $^{139}$La($n,\gamma$) reaction using a guided beam of subthermal (thermal and cold) neutrons incident upon a $^{\rm nat}$La$_{2}$O$_{3}$ target at the Prompt Gamma Activation Analysis facility of the Budapest Research Reactor.  Absolute $^{140}$La cross sections were determined relative to the well-known comparator $^{35}$Cl($n,\gamma$) cross sections from the irradiation of a stoichiometric $^{\rm nat}$LaCl$_{3}$ sample.  The total cross section for radiative thermal neutron-capture on $^{139}$La from the sum of experimentally measured cross sections observed to directly feed the $^{140}$La ground state was determined to be $\sigma_{0}^{\rm expt} = 8.58(50)$~b.  To assess completeness of the decay scheme and as a consistency check, the measured cross sections for transitions feeding the ground state from levels below a critical energy of $E_{c} = 285$~keV were combined with a modeled contribution accounting for ground-state feeding from the quasicontinuum to arrive at a total cross section of $\sigma_{0} = 9.36(74)$~b.  In addition, a neutron-separation energy of $S_{n} = 5161.005(21)$~keV was determined from a least-squares fit of the measured primary $\gamma$-ray energies to the low-lying levels of the $^{140}$La decay scheme.  Furthermore, several nuclear structure improvements are proposed for the decay scheme.  The measured cross-section and separation-energy results are comparable to earlier measurements of these quantities.

\end{abstract}

\pacs{28.20.Np, 27.60.+j, 24.60.Dr, 21.10.Pc}
\maketitle


\section{\label{sec:intro}Introduction}

The nucleus $^{139}$La is an abundant fission product in the $A \approx 143$ region.  It plays a prominent role in the uranium fuel cycle with a cumulative fission-product yield of $>6\%$ in the case of thermal- and fast-neutron induced fission of $^{233,235}$U, and $>5\%$ in fast-neutron induced fission of $^{239}$Pu \cite{crouch:77}.  Together with its direct fission yield, it also occurs in the $\beta^{-}$-decay chain $^{139}$Xe$\rightarrow$$^{139}$Cs$\rightarrow$$^{139}$Ba$\rightarrow$$^{139}$La.  Consequently, neutron-capture cross sections for $^{139}$La provide an important ingredient for nuclear reactor fuel-related applications including fission-product decay-heat calculations and transmutation studies, as well as the development of improved physics models for calculation-based nuclear forensics tools \cite{whitepaper:15}.  Also, the induced $\gamma$-decay activity from $^{140}$La, produced following neutron capture, and its relatively short half life ($T_{1/2} \approx 1.7$~d \cite{nica:07}) make it suitable for isotopic monitoring at reactor facilities, following shutdown, for example \cite{panikkath:17}.  

This study of the $^{139}$La($n,\gamma$) reaction also provides an opportunity to assess the decay-scheme nuclear structure information for the compound nucleus $^{140}$La as well as the role of the photon strength function (PSF) in describing statistical properties of nuclei near the $N=82$ shell closure.  The PSF has profound implications for determination of reaction rates in astrophysical $r$- and $p$-process nucleosynthesis, and for radiation transport calculations to simulate the distribution of emitted $\gamma$ rays where no experimental data are available.  Recently, a strong low-energy enhancement of the PSF, observed for the first time in $^{56,57}$Fe \cite{voinov:04}, was reported in $^{151,153}$Sm \cite{simon:16}.  Notably, however, this low-energy {\sl upbend} is absent approaching $N=82$ in the lighter-mass nuclei $^{148,149}$Sm \cite{siem:02} and $^{144}$Nd \cite{voinov:15}.  In the case of the samarium nuclei, a pronounced change in the measured PSF is interpreted as a possible shape-transitional effect from a modestly-deformed to near-spherical configuration as a single neutron is removed from the system \cite{siem:02}.  Charged-particle reactions have also been carried out recently to probe the PSF in the lanthanum isotopes \cite{kheswa:15,kheswa:17} where a nonzero limit in strength was observed as $E_{\gamma} \rightarrow 0$, similar to that reported for $^{148}$Sm ($N=86$) and $^{144}$Nd ($N=84$).  This work aims to extend our knowledge of the PSF systematics nearing the shell closure with the $N=83$ nucleus $^{140}$La by comparison of our radiative-capture results with those obtained through charged-particle reactions.

\section{\label{sec:exp}Experimental Method and Data Analysis}

Samples composed of natural lanthanum ($^{139}$La: 99.911\% and $^{138}$La: 0.089\% \cite{berglund:11}) compounds, $^{\rm nat}$LaCl$_{3} \cdot 7$H$_{2}$O (hereafter, LaCl$_{3}$) and $^{\rm nat}$La$_{2}$O$_{3}$ (hereafter, La$_{2}$O$_{3}$), were irradiated with thermal and cold neutron beams at the 10-MW Budapest Research Reactor \cite{belgya:97,rosta:97,rosta:02} to measure the radiative neutron-capture $\gamma$-ray production of $^{140}$La.  Four samples were irradiated over various periods at the target station of the Prompt Gamma Activation Analysis (PGAA) facility \cite{revay:04,szentmiklosi:10,belgya:14}: ``thick'' LaCl$_{3}$ (4497.1~mg, 1.7~h); ``thick'' La$_{2}$O$_{3}$ (1037.1~mg, 2.7~h); ``thin'' LaCl$_{3}$ (450~mg, 3.8~h); and ``thin'' La$_{2}$O$_{3}$ (104~mg, 11.0~h).  The thin-sample measurements allow us to assess the effect of self absorption.  The PGAA setup is located $\sim 33.5$~m downstream of the reactor wall at the terminus of a slightly curved, guided neutron beamline.  The guide is comprised of $2\theta_{c}$-supermirror units 0.75~m in length.  Epithermal and higher-energy neutrons have the incorrect wavelength for transmission resulting in a pure beam of thermal neutrons.  A modular aluminium flight tube at the end of the beamline containing a $^{6}$Li-doped polymer to reduce neutron scattering ensures a well-collimated beam at the target position of the PGAA sample chamber \cite{lindstrom:pgaa}.  In this experiment fluxes of $2.3 \times 10^{6}$~n/cm$^{2}$/s (thermal) and $7.75 \times 10^{7}$~n/cm$^{2}$/s (cold) on target for the thick and thin-sample irradiations, respectively, were achieved.

\begin{figure*}[t!]
\begin{center}
\includegraphics[width=1\textwidth]{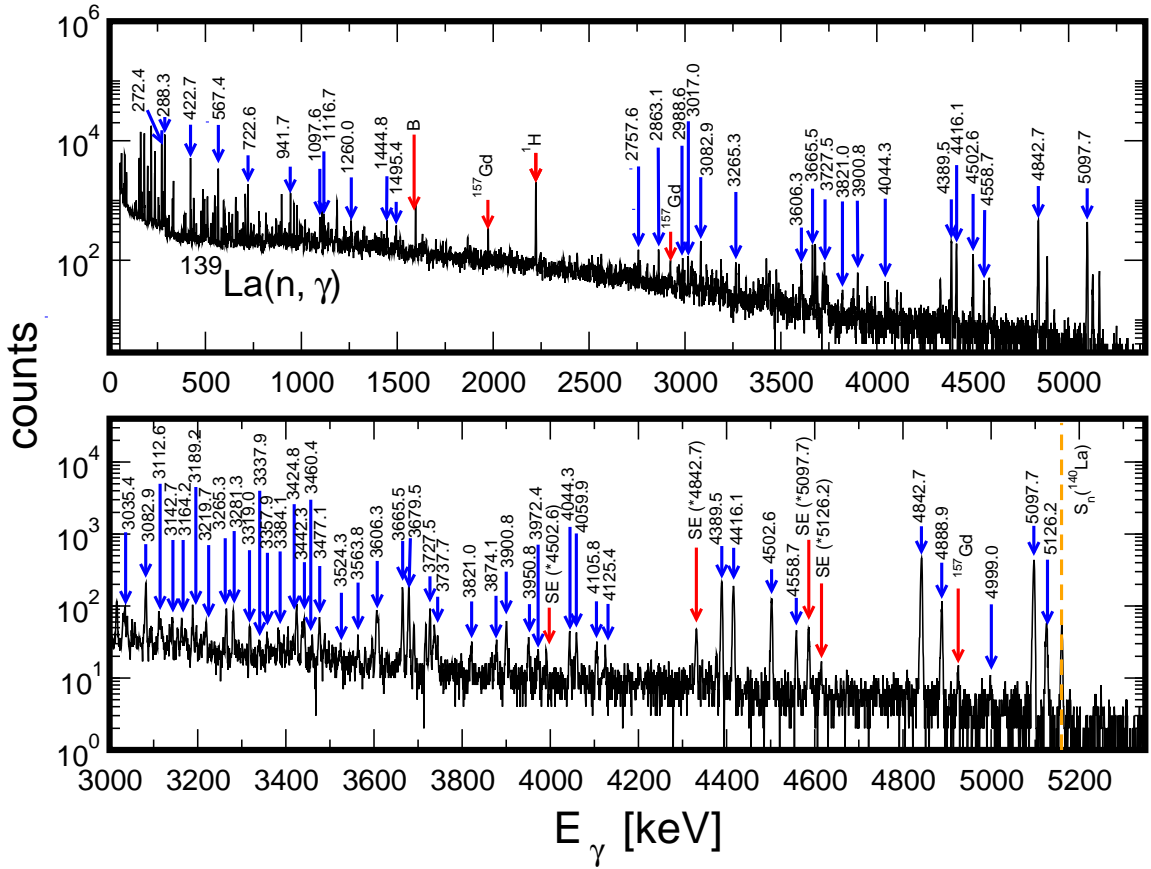}
\end{center}
\caption{(Color online) Prompt $\gamma$-ray energy spectra observed following the $^{139}$La(n,$\gamma$) reaction with the La$_{2}$O$_{3}$ sample.  The upper panel reveals the complete energy range of interest in the capture-$\gamma$ product $^{140}$La.  A strong background line at around 2223~keV from $^{1}$H($n,\gamma$) \cite{revay:pgaa} is clearly visible, as are a few contaminant lines from $^{157}$Gd($n,\gamma$).  In the lower panel, the same spectrum is expanded around $E_{\gamma} = 3000-5350$~keV to highlight many of the intense primary $\gamma$ rays observed in this measurement.  Prominent background lines (B) and single-escape peaks (SE) are labeled as is the position of the neutron-separation energy ($S_{n}$) for $^{140}$La.}
\label{fig:LaOspec}
\end{figure*}

An $n$-type closed-end coaxial high-purity germanium (HPGe) detector is used to monitor $\gamma$ decay at the target-sample position of the PGAA facility.  This detector has an active volume of 144~cm$^{3}$ with a relative efficiency of 27\% at 1332~keV and is surrounded by an annular Compton-suppression shield consisting of eight coaxial bismuth germanate (BGO) segments.  This detection system is mounted within 10-cm-thick lead housing that is itself encased by a $^{6}$Li-doped plastic layer for enhanced $\gamma$-ray and neutron absorption to maintain low-background conditions \cite{szentmiklosi:10}.  The face of the HPGe detector is located 23.5~cm from the target position to minimize peak-summing effects \cite{revay:04} and is oriented at $90^{\circ}$ to the beam direction, while the target sample is held in a thin Teflon bag oriented at $30^{\circ}$ to the beam line.  The BGO segments are operated in anticoincidence mode with respect to the HPGe detector to veto Compton-scattering events and thereby reduce their impact on the observed $\gamma$-ray continuum.  Energy and efficiency calibrations of the HPGe spectrometer were performed using a variety of standard radioactive ($^{133}$Ba, $^{152}$Eu, $^{207}$Bi, $^{226}$Ra, and $^{241}$Am) and reaction sources (deuterated urea: $^{14}$N($n,\gamma$)$^{15}$N and polyvinyl chloride: $^{35}$Cl($n,\gamma$)$^{36}$Cl) for energies below and above 1.5~MeV, respectively.  The nonlinearity and efficiency curves were generated using the $\gamma$-ray spectroscopy software package {\small HYPERMET-PC} \cite{szentmiklosi:10,fazekas:97,revay:01,hypermet}; this program was also used in the offline analysis of the prompt $\gamma$-ray spectra.  Representative prompt spectra from the thick La$_{2}$O$_{3}$ measurement are shown in Fig.~\ref{fig:LaOspec}; several primary and secondary $\gamma$-ray transitions corresponding to the $^{139}$La($n,\gamma$) reaction are clearly visible.

\subsection{\label{sec:expt.A}Cross-section standardization procedure\protect\\}

Two of the lanthanum samples irradiated were primarily used to obtain the set of partial $\gamma$-ray production cross sections ($E_{\gamma} \geq 200$~keV) for the $^{140}$La compound presented in this work: the thick LaCl$_{3}$ and the thick La$_{2}$O$_{3}$ samples.  The former was used to extract cross sections for several prompt $^{139}$La($n,\gamma$) capture lines relative to those from transitions in the well-known comparator $^{36}$Cl: $^{35}$Cl($n,\gamma$) \cite{molnar:04}.  These cross sections, corresponding to strong, well-resolved transitions covering the observed excitation-energy range in $^{36}$Cl are listed in Table~\ref{tab:standards}.  Adopting the internal-standardization procedure described in Ref.~\cite{revay:03}, and assuming $\gamma$-ray self absorption is negligible above 200~keV (see Sect.~\ref{sec:expt.B}), it can be shown (see, e.g., Ref.~\cite{hurst:15}):
\begin{equation}
\label{eq:standard}
\frac{\sigma_{\gamma,x}}{\sigma_{\gamma,c}} = \frac{n_{x}}{n_{c}} \cdot \frac{A_{\gamma,x}/\epsilon(E_{\gamma,x})}{A_{\gamma,c}/\epsilon(E_{\gamma,c})},
\end{equation}
where $\sigma_{\gamma,x}$ and $\sigma_{\gamma,c}$ denote the partial $\gamma$-ray production cross sections for the unknown ($x$) and comparator ($c$) $\gamma$-ray lines, respectively.  Similarly, $A_{\gamma,x}$ and $A_{\gamma,c}$ are the measured peak areas of the unknown and comparator $\gamma$ rays, respectively, $\epsilon(E_{\gamma})$ is the detector efficiency at $\gamma$-ray energies $E_{\gamma,x}$ and $E_{\gamma,c}$, and $(n_{x}/n_{c})$ accounts for the stoichiometry of the irradiated sample.  This expression holds for so-called regular ``$1/v$ nuclides'' where $\sigma_{\gamma}(E_{n}) \propto 1/v$; $\forall~ E_{n} \leq 25.3$~meV.  Both $^{35}$Cl and $^{139}$La fall into this category as they have Westcott $g$ factors that deviate from unity by less than $1\%$ at $T=293$~K and no correction is needed for the neutron-beam temperature in these cases \cite{choi:iaea}.  

\begin{figure}[ht]
  \includegraphics[angle=0,width=\linewidth]{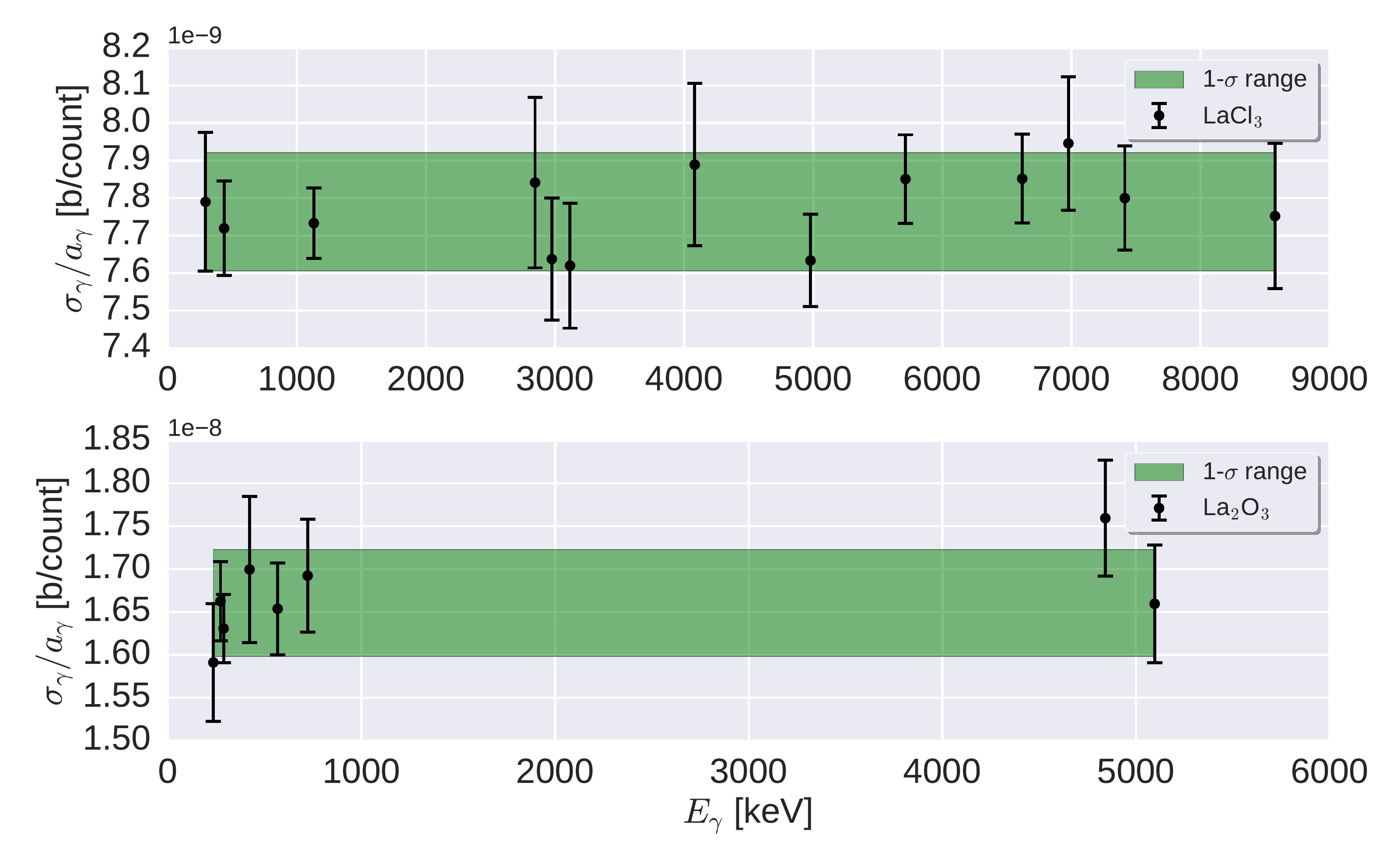}
  \caption{\label{fig:standards} (Color online) Normalization $\gamma$ rays used to standardize the LaCl$_{3}$ and La$_{2}$O$_{3}$ spectra.  The expectation value of the ratio $\sigma_{\gamma}/a_{\gamma}$ is indicated by the shaded 1-$\sigma$ range on each plot.  The standardized cross sections used in these plots are listed in Table~\ref{tab:standards} and were deduced from the LaCl$_{3}$ sample.}
\end{figure}

Equation~(\ref{eq:standard}) shows that the cross section for any given transition should be directly proportional to its efficiency-corrected peak area, i.e., $\sigma_{\gamma} \propto a_{\gamma}$ where $a_{\gamma} = A_{\gamma}/\epsilon(E_{\gamma})$ and, thus, the ratio $\sigma_{\gamma}/a_{\gamma}$ should be constant.  These ratios have been measured in the thick LaCl$_{3}$ spectrum for all $^{36}$Cl comparator lines in Table~\ref{tab:standards} and the resulting plot of Fig.~\ref{fig:standards} reveals statistical consistency for all values, implying self absorption is not an issue at these $\gamma$-ray energies.  We have averaged these results, indicated by the shaded region on the plot, to represent the expectation value of this constant ratio $\langle N \rangle$.  Accordingly, standardized cross sections for $^{\rm nat}$La($n,\gamma$) can then be determined using this normalization factor as
\begin{equation}
\label{eq:standard2}
\sigma_{\gamma,x} = \frac{n_{x}}{n_{c}} a_{\gamma,x}  \langle N \rangle.
\end{equation}
The known $1({\rm La}): 3({\rm Cl})$ stoichiometry of the irradiated LaCl$_{3}$ sample implies $(n_{x}/n_{c}) = 1/3$.  The $^{\rm nat}$La($n,\gamma$) cross sections obtained using Eq.~(\ref{eq:standard2}) yield insignificant changes upon correction for isotopic abundance to arrive at isotopic $^{139}$La($n,\gamma$) cross sections.  There was no obvious contamination from capture on $^{138}$La($\sigma_{0} = 51(5)$~mb \cite{mughabghab:06} assuming the reported 0.089\% abundance \cite{berglund:11}).  However, the complexity of the LaCl$_{3}$ spectrum renders it difficult to unambiguously resolve all lanthanum capture-$\gamma$ lines due to interference from chlorine lines.  The subset of strong prompt $^{140}$La $\gamma$-ray transitions listed in Table~\ref{tab:standards} were selected to cover the observed energy range and used as standards for normalizing the intensities of all prompt $\gamma$ rays $>200$~keV measured in the much cleaner La$_{2}$O$_{3}$ spectrum.  This spectrum is shown in Fig.~\ref{fig:LaOspec}.  Normalization methods for lower-energy $\gamma$ rays are discussed in the following Sect.~\ref{sec:expt.B}.

Because we are only interested in $\gamma$ rays belonging to lanthanum from the La$_{2}$O$_{3}$ measurement, and lanthanum $\gamma$ rays are also being used for comparative purposes here, stoichiometry considerations are redundant, i.e., $n_{x}/n_{c} = 1$.  Furthermore, any oxygen capture is insignificant owing to its much smaller total radiative neutron-capture cross section; the oxygen isotopes have values five orders of magnitude lower \cite{firestone:16} than the currently adopted value for $^{139}$La($n,\gamma$) \cite{mughabghab:06}.  In addition, the oxygen capture-$\gamma$ spectrum is known to be rather sparse and weak \cite{firestone:16, revay:pgaa}, posing an unlikely source of significant contamination and no oxygen-capture lines were observed.  Thus, using Eqs.~(\ref{eq:standard}) and (\ref{eq:standard2}) together with the set of $^{140}$La comparator $\gamma$-ray lines listed in Table~\ref{tab:standards}, a suitable cross-section normalization (Fig.~\ref{fig:standards}; lower panel) was determined for all prompt $^{139}$La($n,\gamma$) lines measured in this work.

\begin{figure}[ht]
  \includegraphics[angle=0,width=\linewidth]{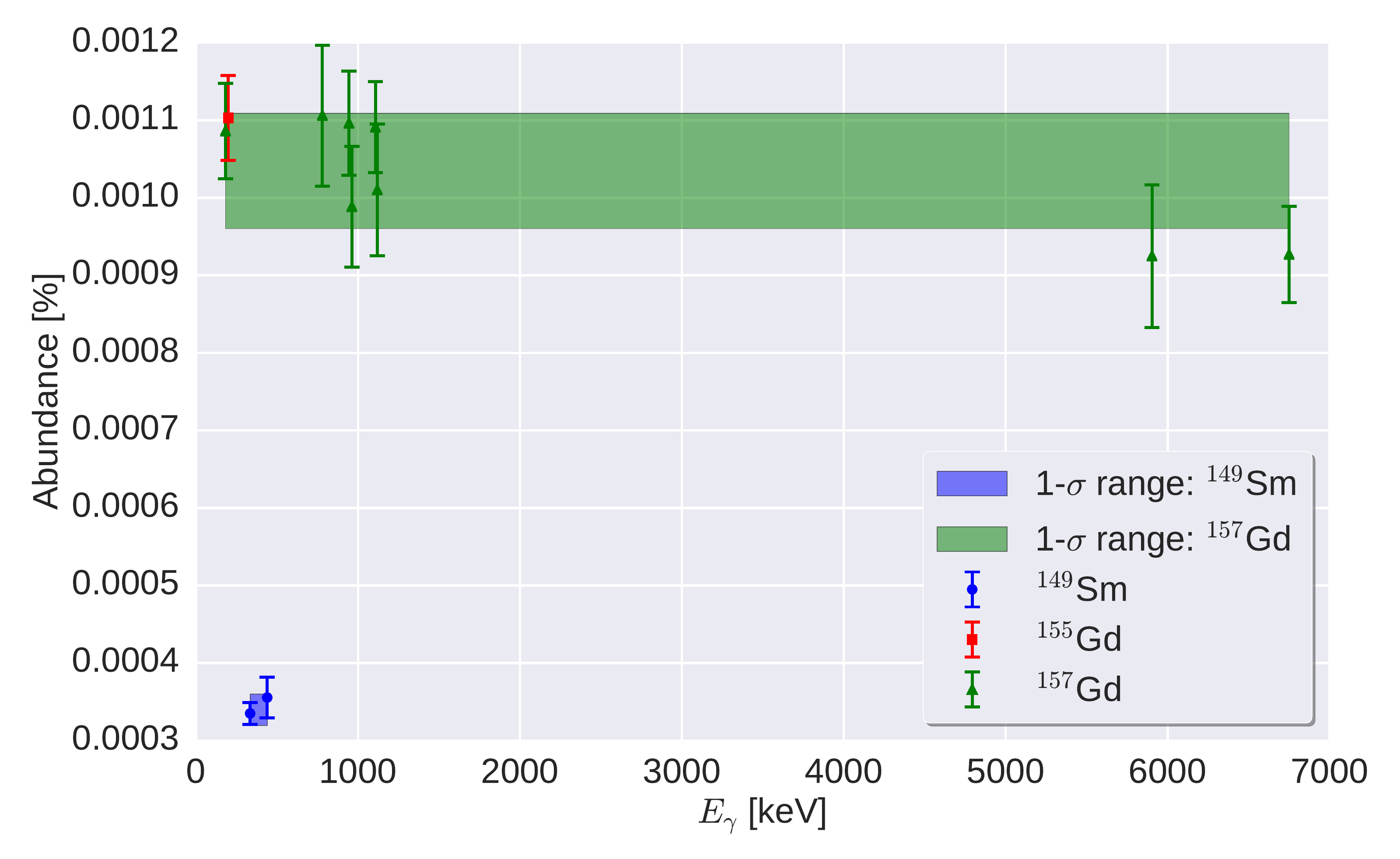}
  \caption{\label{fig:abundances} (Color online) Measured isotopic abundances of $^{149}$Sm and $^{155,157}$Gd contaminants in the La$_{2}$O$_{3}$ sample.}
\end{figure}

Lanthanide targets that have not been isotopically enriched typically suffer from low levels of contamination from other rare-earth isotopes.  Indeed, a few strong lines from $^{155,157}$Gd and $^{149}$Sm were observed in the prompt $(n,\gamma$) spectra.  By comparing the standardized partial $\gamma$-ray cross sections, deduced using Eq.~(\ref{eq:standard2}), for transitions in these isotopes to their known cross sections \cite{choi:14, revay:pgaa}, we have established isotopic compositions of $1.10(5)\times10^{-3}\%$, $1.03(7)\times10^{-3}\%$, and $3.4(2)\times10^{-4}\%$ for $^{155}$Gd, $^{157}$Gd, and $^{149}$Sm, respectively, in the $^{\rm nat}$La$_{2}$O$_{3}$ sample.  Figure~\ref{fig:abundances} shows the transitions used to determine these abundances.  The isotopes $^{155}$Gd and $^{157}$Gd occur naturally together in roughly equal proportions: 14.8\% and 15.7\% \cite{berglund:11}, respectively.  There is no reason to expect this ratio to change in a natural sample and all measured lines are in accordance with it.  Only transitions from $^{157}$Gd($n,\gamma$) overlapped with lines of interest in $^{139}$La($n,\gamma$).  These transitions were clearly identified and accounted for in determining the partial $\gamma$-ray cross sections presented in Sect.~\ref{sec:results}.

\subsection{\label{sec:expt.B}Low-energy $\gamma$ rays and photon attenuation\protect\\}

Low-energy transitions ($E_{\gamma} \lesssim 200$~keV) can be significantly attenuated in high-density materials, e.g., tungsten \cite{hurst:14} and rhenium \cite{matters:16}, requiring significant corrections for photon attenuation \cite{hurst:15B}.  
Because the thick samples used in this measurement are rather bulky, however, to ensure low-energy $\gamma$ rays are treated appropriately for photon attenuation, two additional measurements using thin samples of the exact same lanthanum compounds were subsequently performed.  The thin-sample measurements were again standardized using the procedure outlined in Sect.~\ref{sec:expt.A} to obtain a set of partial $\gamma$-ray production cross sections for the $^{139}$La($n,\gamma$) reaction; all reference cross sections listed in Table~\ref{tab:standards} are consistent with the thin-sample measurements.  Unfortunately, many transitions in the low-energy region of the spectrum are still difficult to obtain precision measurements for due to overlapping peaks and $\gamma$-ray self absorption.  To address this issue, we have exploited the $\gamma$-ray data of Meyer \textit{et al}. \cite{meyer:90} which reports absolute intensities corrected for self absorption for the low-energy spectrum $E_{\gamma} < 70$~keV in $^{140}$La.  Here, we compare our normalized cross sections for the 45.9-, 49.8-, and 56.3-keV $\gamma$ rays deexciting the 318.2-, 322.0-, and 658.3-keV levels, respectively, to the absolute intensities per 100 neutron captures ($I_{\gamma}/100n$) in Ref.~\cite{meyer:90}.  Although these are very low-energy transitions, our cross sections are reliable because we were able to normalize to intense well-resolved higher-energy transitions (where self absorption in the target is negligible) deexciting the same levels according to their known branching ratios \cite{nica:07} (see Table~\ref{tab:gammas}).  Based on consistency between the $I_{\gamma}/100n$ data \cite{meyer:90} relative to our cross sections for these transitions, we could, thus, convert $I_{\gamma}/100n$ $\gamma$-ray measurements to absolute standardized cross sections for all other low-energy transitions from Ref.~\cite{meyer:90} where $E_{\gamma} < 70$~keV.  All other cross sections were obtained from the standardization procedure.  We also performed a consistency check in the $\gamma$-ray energy region around $E_{\gamma} = 100-200$ keV, whereupon our deduced thin-sample standardized cross sections compare well with the absorption-corrected absolute intensities per $100n$ reported in the earlier $^{139}$La($n,\gamma)$ work by Jurney \emph{et al}. \cite{jurney:70} that covers a broader interval of $\gamma$-ray energies up to 1500~keV.  The observed consistency implies self absorption is not a prevalent issue in this energy region in the thin-sample measurements.

\begin{table}[t]
  \centering
  \caption{\label{tab:standards} Comparator $\gamma$-ray lines in $^{36}$Cl and $^{140}$La compounds together with their corresponding standardized partial $\gamma$-ray production cross sections obtained from the LaCl$_{3}$ sample.  The $^{139}$La($n,\gamma$) values were then adopted as normalization cross sections for the La$_{2}$O$_{3}$ measurement.} 
  \begin{tabular*}{0.48\textwidth}{@{\extracolsep{\fill}}ccc@{}}
    \hline\hline
    Source & $E_{\gamma}$ (keV) & $\sigma_{\gamma,c}$ (b) \\
    \hline  
    $^{36}$Cl: $^{35}$Cl($n,\gamma$) & 292.2  	& 0.0893(10) \\
    $^{36}$Cl: $^{35}$Cl($n,\gamma$) & 436.2  	& 0.3093(20) \\
    $^{36}$Cl: $^{35}$Cl($n,\gamma$) & 1131.3 	& 0.6262(33) \\
    $^{36}$Cl: $^{35}$Cl($n,\gamma$) & 2845.5  	& 0.3495(26) \\
    $^{36}$Cl: $^{35}$Cl($n,\gamma$) & 2975.2 	& 0.3765(43) \\
    $^{36}$Cl: $^{35}$Cl($n,\gamma$) & 3116.0 	& 0.2975(26) \\
    $^{36}$Cl: $^{35}$Cl($n,\gamma$) & 4082.7 	& 0.2629(49) \\
    $^{36}$Cl: $^{35}$Cl($n,\gamma$) & 4979.8 	& 1.2320(99) \\
    $^{36}$Cl: $^{35}$Cl($n,\gamma$) & 5715.2 	& 1.818(16) \\
    $^{36}$Cl: $^{35}$Cl($n,\gamma$) & 6619.6	& 2.530(23) \\
    $^{36}$Cl: $^{35}$Cl($n,\gamma$) & 6977.8 	& 0.7412(99) \\
    $^{36}$Cl: $^{35}$Cl($n,\gamma$) & 7414.0 	& 3.291(46) \\
    $^{36}$Cl: $^{35}$Cl($n,\gamma$) & 8578.6  	& 0.883(13) \\
    $^{140}$La: $^{139}$La($n,\gamma$) & 235.8    & 0.1017(35) \\
    $^{140}$La: $^{139}$La($n,\gamma$) & 272.4    & 0.494(12) \\
    $^{140}$La: $^{139}$La($n,\gamma$) & 288.3    & 0.698(16) \\
    $^{140}$La: $^{139}$La($n,\gamma$) & 422.7    & 0.364(12) \\
    $^{140}$La: $^{139}$La($n,\gamma$) & 567.4    & 0.3318(97) \\
    $^{140}$La: $^{139}$La($n,\gamma$) & 722.4    & 0.2247(78) \\
    $^{140}$La: $^{139}$La($n,\gamma$) & 4842.3   & 0.667(21) \\
    $^{140}$La: $^{139}$La($n,\gamma$) & 5097.3   & 0.650(22) \\
    \hline\hline
    \end{tabular*}
\end{table}

\begin{figure}[t]
  \includegraphics[angle=0,width=\linewidth]{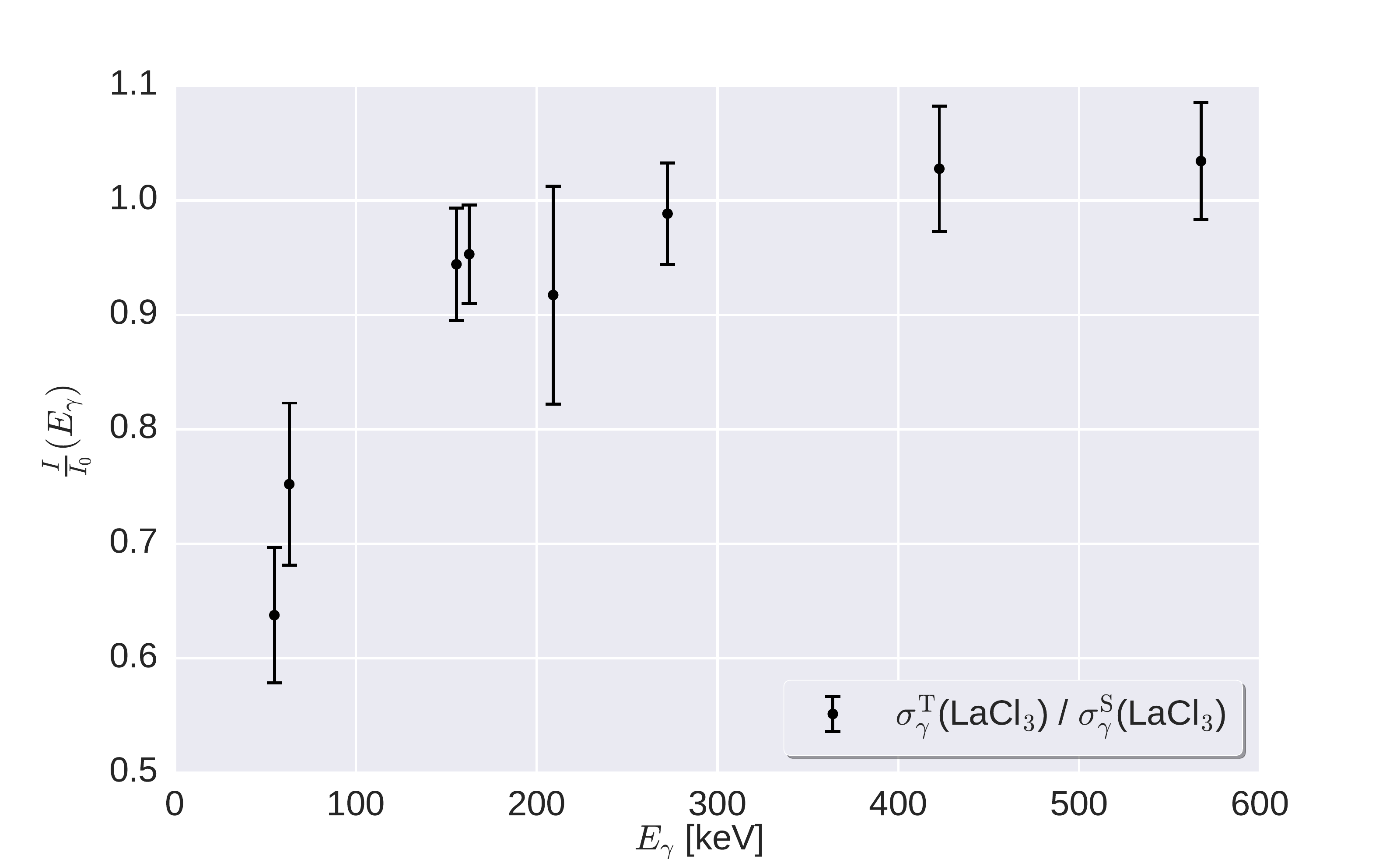}
  \caption{\label{fig:attenuation} (Color online) Ratios of standardized partial $\gamma$-ray production cross sections for clean prompt $^{139}$La($n,\gamma$) transitions measured in the thick LaCl$_{3}$ sample (T) relative to the same transitions in the thin reference LaCl$_{3}$ sample (S).  The ratios were measured at the following $\gamma$-ray energies: 54.9, 63.2, 155.6, 162.6, 209.2, 272.4, 422.7, and 567.4~keV.}
\end{figure}

The standardized cross sections obtained from the thin-sample measurements and renormalized absolute intensity data may then be compared to those from the thick-sample measurements to assess the effect of $\gamma$-ray self absorption.  The experimental photon attenuation at a given $\gamma$-ray energy (in a thick sample) with measured intensity $I_{\gamma}(E_{\gamma})$, may be determined by comparing partial $\gamma$-ray production cross sections from an optically-thin reference sample ($\sigma_{\gamma}^{\rm S}$) to those from the thick-target sample ($\sigma_{\gamma}^{\rm T}$):
\begin{equation}
\label{eq:attenuation}
\left( \frac{I(E_{\gamma})}{I_{0}} \right)_{\rm expt} = \frac{\sigma_{\gamma}^{\rm T}}{\sigma_{\gamma}^{\rm S}},
\end{equation}
where $I_{0}$ is the unattenuated intensity produced in the sample.  In the absence of any attenuation, this ratio should be around unity.  In the event of significant $\gamma$-ray absorption $\sigma_{\gamma}^{\rm T} < \sigma_{\gamma}^{\rm S}$.  Figure~\ref{fig:attenuation} shows the ratios of standardized cross sections for several clean transitions below 600~keV from the thick hydrated LaCl$_{3}\cdot7$H$_{2}$O ($\rho = 2.23$~g/cm$^{3}$ \cite{LaCl3_7H2O:density}) sample ($\sigma_{\gamma}^{\rm T}$) relative to the corresponding standardized cross sections from the thin sample ($\sigma_{\gamma}^{\rm S}$) of the same compound.  The effect of self absorption is clear for $\gamma$-ray transitions with $E_{\gamma} < 200$~keV, while ratios deduced with Eq.~(\ref{eq:attenuation}) for all transitions above this energy are consistent with unity.  Similar results were obtained with the La$_{2}$O$_{3}$ ($\rho = 6.51$~g/cm$^{3}$) sample.

\begin{figure*}[t]
  \includegraphics[angle=0,width=\linewidth]{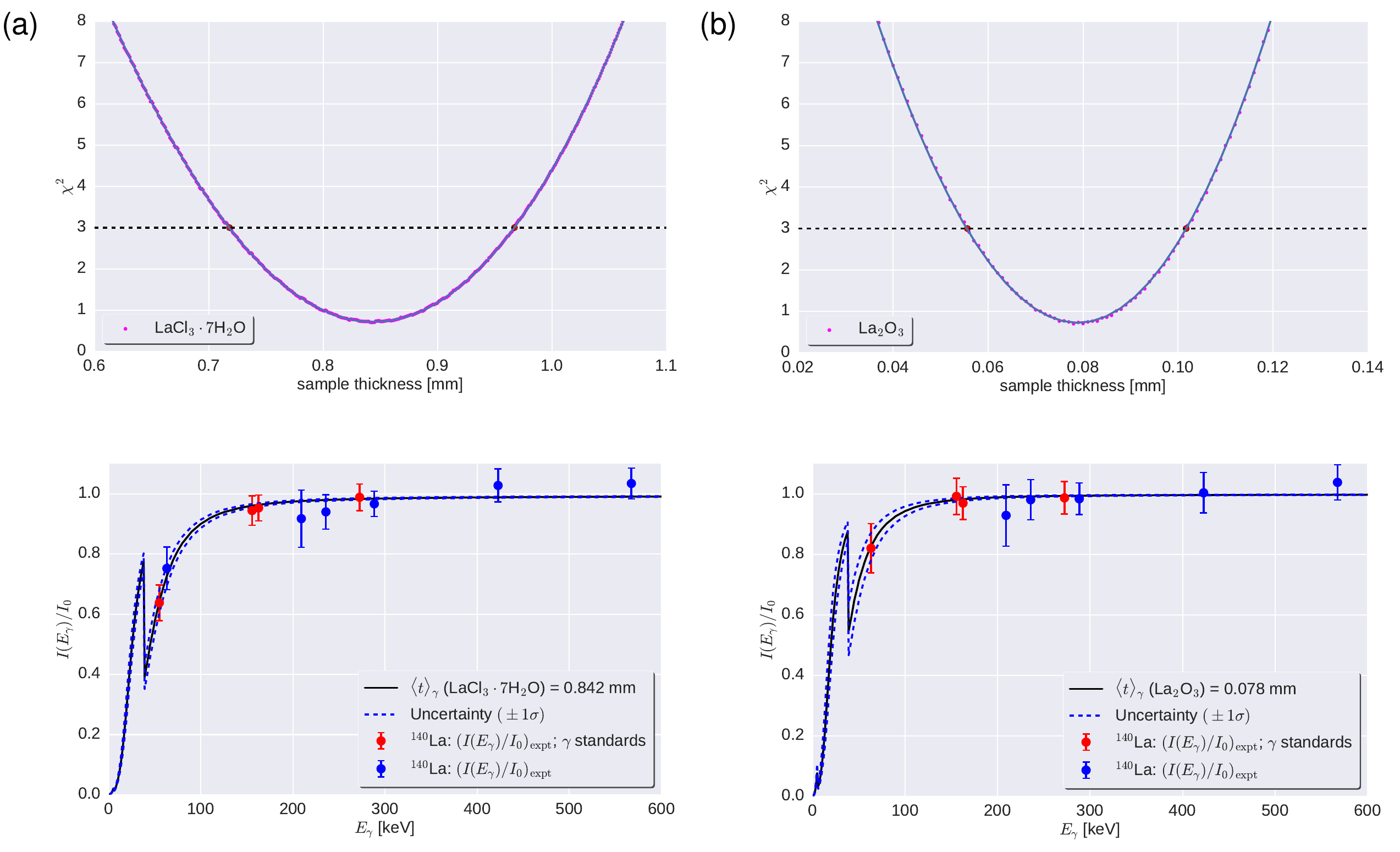}
  \caption{\label{fig:chi2} (Color online) Upper panels: plot of the global $\chi^{2}$ values as a function of the target thickness for (a) LaCl$_{3}\cdot 7$H$_{2}$O and (b) La$_{2}$O$_{3}$.  The dashed line drawn at $\chi^{2}_{\rm min} + 2.3$ defines the $1\sigma$ range of acceptable $t$ values for two adjustable parameters \cite{beringer:12}.  Lower panels: plot of the calculated attenuation factor at the deduced effective sample thicknesses of (a) $\langle t \rangle = 0.842 (^{+125}_{-124})$~mm for LaCl$_{3}\cdot 7$H$_{2}$O, and (b) $\langle t \rangle = 0.078 (^{+24}_{-22})$~mm for La$_{2}$O$_{3}$.  The experimentally measured attenuation factors (Eq.~(\ref{eq:attenuation})) for a subset of strong well-resolved $\gamma$ rays are indicated on each plot.  The red data points indicate the $\gamma$-ray standards used to constrain the fits: 54.9-keV (LaCl$_{3}\cdot 7$H$_{2}$O only); 63.2-keV (La$_{2}$O$_{3}$ only); 155.6-keV; 162.6-keV; and 272.4-keV transitions.  Also shown for comparison are the 209.2-, 235.8-, 288.3-, 422.7-, and 567.4-keV transitions (blue data points) that were not used to constrain the fitting procedure.}
\end{figure*}

In order to validate our results, we calculated the effective sample thickness $t$ consistent with the observed photon attenuation for each of the thick lanthanum samples according to the prescription of Ref.~\cite{hurst:15B}.  Using this methodology, the attenuation is determined by integrating the exponential-attenuation law $I_{\gamma}/I_{0} = \exp(-\mu_{\gamma}x)$ over the sample depth $x$:
\begin{equation}
\frac{I_{\gamma}}{I_{0}} \int\limits^{x=t}_{x=0} dx = \int\limits^{x=t}_{x=0} \exp\left(\frac{-\rho\left(\frac{\mu_{\gamma}}{\rho}\right)_{E_{\gamma}}x}{\cos \theta}\right) dx, 
\label{eq:int1}
\end{equation}
to arrive at the $E_{\gamma}$- and $t$-dependent expression for the photon attenuation given by
\begin{equation}
\frac{I_{\gamma}(E_{\gamma},t)}{I_{0}}=\frac{\cos \theta}{\rho\left(\frac{\mu_{\gamma}}{\rho}\right)_{E_{\gamma}}t}\left[1-\exp\left(\frac{-\rho\left(\frac{\mu_{\gamma}}{\rho}\right)_{E_{\gamma}}t}{\cos \theta}\right) \right].
\label{eq:int2}
\end{equation}
Here, $(\mu_{\gamma}/\rho)_{E_{\gamma}}$ represents the $\gamma$-ray mass-attenuation coefficient at a given $\gamma$-ray energy $E_{\gamma}$ and is sourced from the {\small XMUDAT} database \cite{xmudat:98}, $\rho$ is the density of the target sample, and $\theta=30^{\circ}$ is the angle of the sample face relative to the detector face in this experiment \cite{revay:09}.  From Eqs.~(\ref{eq:standard}) and (\ref{eq:standard2}), it is clear that the measured peak areas, after correcting for photon attenuation and detector efficiency, should be directly proportional to their corresponding reference cross sections and, therefore, the following relation should hold for all values of $E_{\gamma}$:
\begin{equation}
\frac{\sigma^{\rm S}}{a_{\gamma}} \cdot \frac{I_{\gamma}(E_{\gamma},t)}{I_{0}} = C.
\label{eq:const}
\end{equation}
The effective thickness $t$ for the thick-sample lanthanum compounds was then varied until the constant $C$ in Eq.~(\ref{eq:const}) converged to a unique value for an adopted set of $\gamma$ rays.  For each sample, we performed a $\chi^{2}$ minimization \cite{hurst:15B} using four $\gamma$-ray data points and treated both $t$ and the correlation coefficient as adjustable parameters to minimize the $\chi^{2}$ function leaving two degrees of freedom (${\rm ndf}=2$).  The $\gamma$-ray energies and corresponding standardized partial $\gamma$-ray cross sections used in the minimization procedure are listed in Table~\ref{tab:mindata}.  For a 2-parameter adjustment, the $1\sigma$-uncertainty band is defined by $\chi^{2}_{\rm min} + 2.3$ \cite{beringer:12}.  The resulting $\chi^{2}$ plots are shown in Fig.~\ref{fig:chi2}, revealing expectation values of $\langle t \rangle = 0.078 (\substack{+24 \\ -22})$~mm for La$_{2}$O$_{3}$ and $\langle t \rangle = 0.842 (\substack{+125 \\ -124})$~mm for LaCl$_{3}\cdot 7$H$_{2}$O, according to the observed minima and extracted uncertainty range in each case.  The associated attenuation curves at the deduced thicknesses are also shown in the lower panels of Fig.~\ref{fig:chi2}; the experimentally-deduced photon-attenuation ratios given by Eq.~(\ref{eq:attenuation}) for a subset of well-defined $\gamma$-ray transitions (covering the energy region of interest) are plotted for comparison, and the $\gamma$-ray standards used to constrain each fit (selected from Table~\ref{tab:mindata}) are indicated.  These plots reveal statistical consistency between the calculated photon attenuation at the deduced sample thicknesses and the experimental ratios given by Eq.~(\ref{eq:attenuation}).  

\begin{table}[b]
  \centering
  \caption{\label{tab:mindata} Low-energy standardized $\gamma$-ray cross sections used to deduce the effective thicknesses for the thick LaCl$_{3}\cdot 7$H$_{2}$O and La$_{2}$O$_{3}$ samples.  See text for details.} 
  \begin{tabular*}{0.48\textwidth}{@{\extracolsep{\fill}}lc@{}}
    \hline\hline
    $E_{\gamma}$ (keV) & $\sigma_{\gamma}$ (b) \\
    \hline  
    54.9   & 0.138(12)\\
    63.2   & 0.216(19)\\
    155.6  & 0.1872(80)\\
    162.6  & 0.475(18)\\
    272.4  & 0.499(19)\\
    \hline\hline
    \end{tabular*}
\end{table}

Based on our analysis of the thick- and thin-sample $\gamma$-ray data, three distinct energy regions were categorized in order to discriminate and obtain the complete set of partial $\gamma$-ray production cross sections for $^{140}$La: (i) Low-energy transitions corresponding to $E_{\gamma} \lesssim 70$~keV were obtained from the renormalized $I_{\gamma}/100n$ data of Ref.~\cite{meyer:90} (in particular those deexciting levels below 103.8~keV) unless stated otherwise.  (ii) Moderately low-energy transitions $E_{\gamma} < 200$~keV were obtained from the standardization of the thin-sample measurements.  (iii) All other higher-energy cross sections, where $\gamma$-ray self absorption is shown to be statistically insignificant, were obtained from the standardization of the thick-sample La$_{2}$O$_{3}$ measurement.

\section{\label{sec:theory}Statistical-model calculations\protect\\}

The Monte Carlo statistical-decay code {\small DICEBOX} \cite{becvar:98} was used to simulate the thermal neutron-capture $\gamma$-ray cascades for the compound nucleus $^{140}$La.  Below a certain cutoff energy, the critical energy $E_{c}$, the code takes level energies, spins, parities, and $\gamma$-ray branching information from available experimental data.  Internal conversion is accounted for using coefficients generated from the Band-Raman Internal Conversion Calculator ({\small BRICC}) \cite{kibedi:08}.  In addition, experimental data are also used for intensities of primary $\gamma$ rays deexciting the neutron-capturing state to levels below $E_{c}$.

Above $E_{c}$, {\small DICEBOX} generates a random discrete set of levels using a nuclear level density (LD) model $\rho(E,J,\pi)$.  {\small DICEBOX} then uses photon strength function (PSF) models $f^{(XL)}(E_{\gamma})$ for $E1$, $M1$, and $E2$ transitions to generate a set of transition widths $\Gamma_{if}$ from each state to all other states below, where $i$ and $f$ denote the initial and final levels, respectively, involved in a transition.  The calculated widths fluctuate according to a Porter-Thomas \cite{porter:56} distribution:
\begin{equation}
  P(x)=\frac{1}{\sqrt{2\pi x}}e^{-x/2}, 
  \label{eq:PT}
\end{equation}
where $x= \Gamma_{if}^{XL}/\langle \Gamma_{if} \rangle$, and the expectation value is given by
\begin{equation}
  \langle \Gamma_{if} \rangle = \frac{f^{(XL)}(E_{\gamma}) \cdot E_{\gamma}^{(2L+1)}}{\rho(E_{i}, J_{i}, \pi_{i})}. 
  \label{eq:partGamma}
\end{equation}
Individual levels and partial radiation widths are generated randomly; a complete set of these quantities for a given decay-scheme simulation is referred to as a nuclear realization.

Because of the different decay paths involved, the calculated quantities fluctuate among different nuclear realizations even for a given PSF and LD combination.  The statistical variation of the decay-related observables due to the fluctuation properties of the simulated quantities can be represented as the associated uncertainty.  

The code predictions used in this work were the population of levels below $E_{c}$, and the total radiative width of the capturing state.  To obtain adequate statistical variation in the simulated quantities, we performed 50 separate nuclear realizations, each with 100,000 capture-state $\gamma$-ray cascades, for each combination of PSF and LD.

\subsection{\label{sec:theory.1}Determination of the total radiative thermal-neutron capture cross section\protect\\}

The total radiative thermal neutron-capture cross section $\sigma_{0}$ is determined as the sum of contributions from:
\begin{enumerate}[(i)]
\item the sum of observed \textit{experimental} partial $\gamma$-ray production cross sections feeding the ground state from all levels below $E_{c}$ and directly from the capturing state ($\sum_{i=0}^{n}\sigma_{\gamma i0}^{{\rm expt}}$);
\item the \textit{simulated} fraction of transitions feeding the ground state from the quasicontinuum ($P_{0}$), i.e., within the excitation-energy range $E_{c} < E < S_{n}$.
\end{enumerate}
Combining these two quantities allows for the total cross section to be rewritten as
\begin{equation}
  \sigma_{0} = \frac{\sum\limits_{i=0}^{n}\sigma_{\gamma i0}^{{\rm expt}}(1+\alpha_{i0})}{1-P_{0}}, 
  \label{eq:csform}
\end{equation}
where $\alpha_{i0}$ is the internal conversion coefficient for a transition from level $i$ to the ground state.  In reality, the sum of observed ground-state transitions should be smaller than the determined cross section because of the unknown fraction of missing intensity above $E_{c}$.

\begin{table}[t]
  \caption{\label{tab:vonEgidy} Parametrizations for the CTF ($T$ and $E_{0}$) and BSFG ($a$ and $E_{1}$) LD models used in the statistical-model calculations to simulate capture-$\gamma$ cascades in the $^{140}$La compound.  Adopted parameters corresponded to the listed mean values.}
  \begin{tabular*}{0.48\textwidth}{@{\extracolsep{\fill}}lcccc@{}}
    \hline\hline
    Source & $T$ (MeV) & $E_{0}$ (MeV) & $a$ (MeV$^{-1}$) & $E_{1}$ (MeV) \\
    \hline  
    Ref.~\cite{vonegidy:05} & 0.71(5) & $-1.91(38)$ & 13.52(40) & $-1.20(19)$ \\
    Ref.~\cite{vonegidy:09} & 0.69(5) & $-1.79(37)$ & 12.32(38) & $-1.17(16)$ \\
    \hline\hline
  \end{tabular*}
\end{table}

\subsection{\label{sec:theory.2}Level densities\protect\\}

The nuclear LD dependence on excitation energy $E$, spin $J$, and parity $\pi$ in the adopted models is assumed to have a separable form:
\begin{equation}
  \rho(E,J,\pi) = \rho(E)f(J)\pi(E).
  \label{eq:ld1}
\end{equation}
Here, $\rho(E)$ denotes the total density of nuclear levels as a function of energy.  The adopted spin-distribution factor $f(E)$ is typically expressed using the formula \cite{ericson:60}
\begin{equation}
  f(J) = \frac{2J+1}{2\sigma_{c}^{2}} \exp \left( - \frac{J(J+1/2)}{2\sigma_{c}^{2}} \right),
  \label{eq:ld2}
\end{equation}
where $\sigma_{c}$ is a spin-cutoff parameter and $\pi(E)$ describes the parity dependence as a function of excitation energy.  In this study, we considered two different LD models of the form $\rho(E,J) = \rho(E)f(J)$: the constant-temperature formula (CTF) \cite{ericson:59,gilbert:65} and the back-shifted Fermi gas (BSFG) \cite{gilbert:65,newton:56} models.  The adopted LD models and their parametrizations, determined from fitting low-excitation-energy levels and $s$-wave neutron resonances above the neutron-separation energy $S_{n}$ \cite{vonegidy:05,vonegidy:09}, are briefly described below.  

The CTF model assumes a constant nuclear temperature $T$ throughout the entire excitation-energy range and is given as
\begin{equation}
  \rho(E,J) = \frac{f(J)}{T} \exp \left( \frac{E-E_{0}}{T} \right),
  \label{eq:ctf}
\end{equation}
where $E_{0}$ is the energy-backshift parameter to correct for nucleon pairing, while $T$ may be interpreted as a critical temperature for breaking nucleon pairs.  The adopted parametrizations according to the CTF LD model for $^{140}$La are listed in Table~\ref{tab:vonEgidy}.  A parameter-free constant value is assumed for the spin-cutoff used in $f(J)$ (Eq.~(\ref{eq:ld2})): $\sigma_{c} = 0.98A^{0.29}$ from Refs.~\cite{vonegidy:05,vonegidy:88}.

\begin{table}[t]
  \caption{\label{tab:PSF} Resonance parameters determined for the GDER and GQER used in the statistical-model calculations.  Adopted parameters for the GDER corresponded to the listed mean values.  The GDER parameters were obtained from a fit to the $^{139}$La($\gamma,xn$) data in Ref.~\cite{beil:71} and GQER parameters are from a theoretical global parametrization used to describe isovector-isoscalar vibrations.}
  \begin{tabular*}{0.48\textwidth}{@{\extracolsep{\fill}}lccc@{}}
    \hline\hline
    Resonance & $E_{G}$ (MeV) & $\Gamma_{G}$ (MeV) & $\sigma_{G}$ (mb) \\
    \hline
    GDER & 15.31(2) & 4.70(6) & 335.3(16) \\
    GQER & 12.13 & 4.43 & 3.12 \\
    \hline\hline
  \end{tabular*}
\end{table}

The BSFG model is based on the assumption the nucleus behaves like a two-component noninteracting fermionic fluid and is given by
\begin{equation}
  \rho(E,J) = f(J) \frac{\exp[2\sqrt{a(E-E_{1})}]}{12\sqrt{2}\sigma_{c}a^{1/4}(E-E_{1})^{5/4}},
  \label{eq:bsfg}
\end{equation}
where $E_{1}$ is an excitation-energy backshift to correct for the fermion pairing and $a$ is the shell-model LD parameter.  These parametrizations, as applicable to the BSFG LD model for $^{140}$La, are also listed in Table~\ref{tab:vonEgidy}.  The spin-cutoff parameter adopted for the BSFG model was proposed in Ref.~\cite{zhongfu:91} and has an energy dependence given by
\begin{equation}
  \sigma_{c}^{2} = 0.0146 A^{5/3} \cdot \frac{1+\sqrt{1+4a(E-E_{1})}}{2a}.
  \label{eq:sco}
\end{equation}

The parity-distribution function in Eq.~(\ref{eq:ld1}) $\pi(E)$ denotes the fraction of positive- or negative-parity states as a function of energy.  For the fraction of negative-parity states this implies
\begin{equation}
  \pi(E,\rho_{-}) = \frac{\rho_{-}(E)}{\rho_{-}(E) + \rho_{+}(E)},
  \label{eq:ld3}
\end{equation}
and because parity is conserved, the fraction of positive-parity states is simply $\pi(E,\rho_{+}) = 1 - \pi(E,\rho_{-})$.  An empirical model embodying a Fermi-Dirac functional form proposed by Al-Quraishi \textit{et al}. \cite{quraishi:03} was used to describe the parity dependence in this work:
\begin{equation}
  \pi(E) = \frac{1}{2} \bigg{(} 1 \pm\frac{1}{1+\exp[c(E-\delta_{p})]} \bigg{)},
  \label{eq:parity}
\end{equation}
where the sign of the $\pm$ coefficient is determined according to the parity of the ground state, $c$ is a parity-ratio parameter, and $\delta_{p}$ is an energy-shift correction related to pairing.  For $^{140}$La, the low-lying levels are predominantly characterized by negative parity including its ground state and, therefore, a negative sign is used in Eq.~(\ref{eq:parity}) to describe the parity distribution.  The parametrizations from Ref.~\cite{quraishi:03} for the odd-odd nucleus $^{140}$La were assumed for the statistical-model calculations: $c = 3.0$~MeV$^{-1}$ and $\delta_{p} = 0.0253$~MeV.  We also tested parity-independent LD models which were found to be consistent with our adopted parity-dependent approach.

\subsection{\label{sec:theory.3}Photon strength functions\protect\\}

Initial transitions in capture $\gamma$-ray cascades originating at the neutron-capture state are believed to dominantly have electric dipole ($E1$) character.  These $\gamma$ rays are usually modeled using the low-energy tail of the giant dipole electric resonance (GDER).  In the region well above the neutron-separation energy, the shape of the GDER can be probed through photonuclear ($\gamma,n$) measurements.  At these higher energies, the shape of the resonance can usually be well described with a standard Lorentzian, often referred to as the Brink-Axel (BA) model \cite{brink:55,axel:62}.  For $E_{\gamma} \lesssim S_{n}$ the shape of the $E1$ PSF is not well known, and different extrapolations of the BA model are typically used.  

\begin{figure}[b]
  \includegraphics[width=0.5\textwidth]{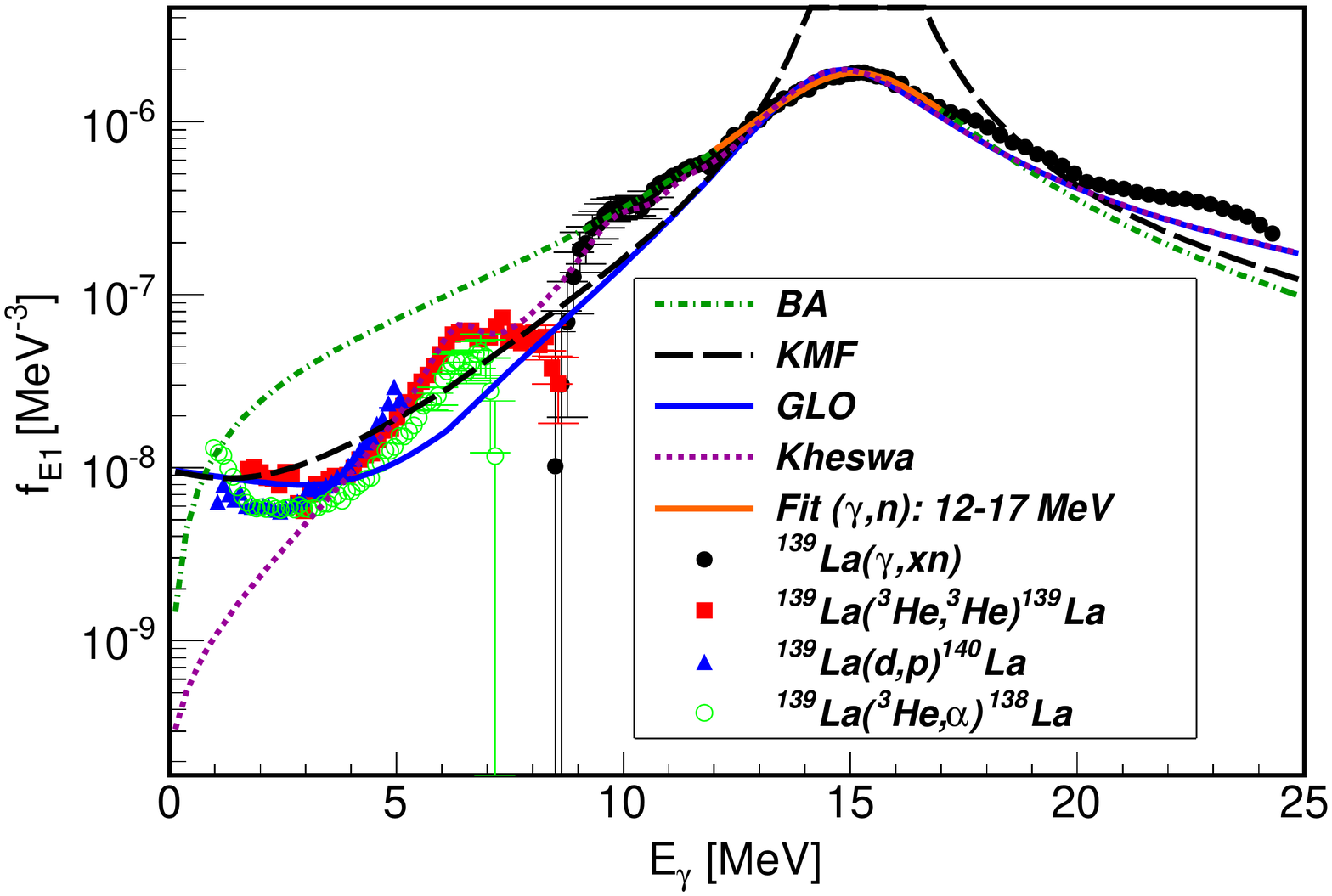}
  \caption{\label{fig:PSF} (Color online) Experimental data overlaid with BA \cite{brink:55,axel:62}, GLO \cite{kopecky:90}, and KMF \cite{kadmenski:83} models describing the $E1$ PSF.  The photonuclear $^{139}$La($\gamma,xn$) data (black circles) are taken from Ref.~\cite{beil:71}, the $^{139}$La($^{3}$He,$^{3}$He') data (red squares) and the $^{139}$La($^{3}$He,$\alpha$) data (green circles) are both taken from Ref.~\cite{kheswa:15}, and the $^{139}$La($d,p$) data (blue triangles) are taken from Ref.~\cite{kheswa:17}.  The dotted-purple curve labeled as Kheswa corresponds to the PSF described in Ref.~\cite{kheswa:15}.  The orange curve represents a Lorentzian fit to the giant dipole resonance observed over 12-17~MeV used to deduce the adopted parametrizations of the PSF.}
\end{figure}

To parametrize the $E1$ PSF we fit a single-component standard Lorentzian to the nearest-neighboring photonuclear $^{139}$La($\gamma,xn$) data of Beil \textit{et al}. \cite{beil:71}, shown in Fig.~\ref{fig:PSF} in the interval $12 \leq E_{\gamma} \leq 17$~MeV (illustrated with the orange curve).  The fitted-resonant energy ($E_{G}$), width ($\Gamma_{G}$), and cross section ($\sigma_{G}$) results are listed in Table~\ref{tab:PSF}.  Our results compare reasonably well with those in the Reference Input Parameter Library (RIPL) \cite{capote:09}.  Using these parameters in Table~\ref{tab:PSF}, we tested not only the BA model but also other models of the $E1$ PSF, namely: the Kadmenski, Markushev, and Furman (KMF) \cite{kadmenski:83} and the generalized Lorentzian (GLO) \cite{kopecky:90} models.

The Brink-Axel function $f^{(E1)}_{\rm BA}$ for the $E1$ PSF is described by a standard Lorentzian according to
\begin{equation} 
  f^{(E1)}_{\rm BA}(E_{\gamma}) = \frac{1}{3(\pi\hbar c)^{2}} \cdot \frac{\sigma_{G} E_{\gamma} \Gamma^{2}_{G}}{(E^{2}_{\gamma}-E^{2}_{G})^{2}+E^{2}_{\gamma}\Gamma^{2}_{G}},
  \label{eq:ba}
\end{equation}
where the constant $\frac{1}{3(\pi\hbar c)^{2}} = 8.68 \times 10^{-8}$~mb$^{-1}$~MeV$^{-2}$.  As shown in Fig.~\ref{fig:PSF}, the BA model describes the high-energy photonuclear data rather well, but it significantly overestimates the low-energy PSF data obtained using the Oslo Method \cite{kheswa:15,kheswa:17}.

The BA model is dependent on $E_{\gamma}$ alone.  Both the KMF and GLO models, on the other hand, include an additional temperature dependence because they embody a temperature-dependent resonance width given by
\begin{equation}
  \Gamma_{G}(E_{\gamma},\Theta) = \frac{\Gamma_{G}}{E^{2}_{G}}(E_{\gamma}^{2} + 4 \pi^{2} \Theta^{2}),
  \label{eq:temp_width}
\end{equation}
where the nuclear temperature $\Theta$ is a function of the excitation energy of the final level $E_{f}$, such that
\begin{equation}
  \Theta = \sqrt{(E_{f}-\Delta)/a}.
  \label{eq:temp}
\end{equation}
In Eq.~(\ref{eq:temp}), $a$ is the shell-model LD parameter described in Sect.~\ref{sec:theory.2} and the pairing energy for odd-odd nuclei is calculated using $\Delta = -0.5|P_{d}|$, where $P_{d}$ is the deuteron-pairing energy tabulated in Ref.~\cite{vonegidy:05}.  For $^{140}$La, $P_{d} = -2.079$~MeV and, thus, $\Delta = -1.0395$~MeV.

The KMF model for the $E1$ PSF is then given by
\begin{equation}
  f^{(E1)}_{\rm KMF}(E_{\gamma},\Theta) = \frac{1}{3(\pi\hbar c)^{2}} \cdot \frac{\sigma_{G} \Gamma_{G} E_{\gamma} \Gamma_{G}(E_{\gamma},\Theta)}{(E^{2}_{\gamma}-E^{2}_{G})^{2}}.
  \label{eq:kmf}
\end{equation}
As in previous works, e.g., Refs.~\cite{hurst:14,hurst:15,matters:16,krticka:08}, we have set the dimensionless Fermi-liquid parameter $F_{K}$ to a value of 0.7 \cite{kadmenski:83,kopecky:98}.  As shown in Fig.~\ref{fig:PSF}, the KMF model describes the Oslo data (i.e. the region $E_{\gamma} \lesssim S_{n}$) much better than the BA model.

The GLO model, developed by Kopecky and Uhl \cite{kopecky:90}, connects the BA model near the maximum of the GDER with the KMF model at low $\gamma$-ray energy.  This model has the following analytic form
\begin{multline} 
  f^{(E1)}_{\rm GLO}(E_{\gamma}) = \frac{\sigma_{G}\Gamma^{2}_{G}}{3(\pi\hbar c)^{2}} \Big[ F_{K}\frac{4\pi^{2}\Theta^{2}\Gamma_{G}}{E^{5}_{G}} \\+ \frac{E_{\gamma}\Gamma_{G}(E_{\gamma},\Theta)}{(E^{2}_{\gamma}-E^{2}_{G})^{2}+E^{2}_{\gamma}\Gamma^{2}_{G}(E_{\gamma},\Theta)} \Big].
  \label{eq:glo}
\end{multline}
Overall, Fig.~\ref{fig:PSF} shows that the GLO model for the $E1$ PSF best reproduces both the low- and high-energy data.

A single-particle (SP) model \cite{blatt:52,bartholomew:61} was adopted in this work to describe the magnetic dipole $M1$ strength.  Although we varied the SP strength to test its influence on the simulations, Fig.~\ref{fig:PSF} suggests that any additional $M1$ strength is likely to be small because the models for the $E1$ PSF already reproduce the low-energy experimental data adequately (KMF and GLO models), or over predict (BA model).  Ultimately, a value of ${\rm SP} = 1 \times 10^{-9}$~MeV$^{-3}$ was adopted for the $M1$ PSF cf. the adopted value in previous studies of $^{180}$W \cite{hurst:15}, $^{183,185,187}$W \cite{hurst:14} and $^{186}$Re \cite{matters:16}.  A scissors mode \cite{loiudice:78,richter:90} for the $M1$ PSF was also considered in this work but it is expected to be very weak and did not produce any significant differences.  

As a further consideration, we also tested the empirical function developed by Kheswa \textit{et al}. \cite{kheswa:15} to fit the low-energy charged-particle \cite{kheswa:15,kheswa:17} and high-energy photonuclear \cite{beil:71,utsunomiya:06} experimental strength function data for several lanthanum isotopes.  This model combines three standard Lorentzians (BA model) with resonances centered on 6.4 MeV ($\sigma_{G} = 2.9$~mb, $\Gamma_{G} = 1.3$~MeV), 9.9~MeV ($\sigma_{G} = 15$~mb, $\Gamma_{G} = 1.6$~MeV), and 11.4~MeV ($\sigma_{G} = 15$~mb, $\Gamma_{G} = 1.4$~MeV), together with a GLO model centered on a GDER with $E_{G} = 15.6$~MeV, $\sigma_{G} = 336$~mb, and $\Gamma_{G} = 5.6$~MeV.  A fixed nuclear temperature $\Theta = 0.1$~MeV was adopted in the GLO component of this model in place of, and to remove the excitation-energy dependence given by, Eq.~(\ref{eq:temp}).  The total fitting function represented by this model is overlaid with the experimental data labeled ``Kheswa'' in Fig.~\ref{fig:PSF}.

Quadrupole strength is expected to contribute far less than dipole strength.  We have modeled the contribution of the $E2$ PSF based on the giant quadrupole electric resonance (GQER).  This model uses a standard Lorentzian (SLO) with a single resonance to describe the isovector-isoscalar quadrupole vibration, given by
\begin{equation}
  f^{(E2)}_{SLO}(E_{\gamma}) = \frac{1}{5(\pi \hbar c)^{2}} \cdot \frac{\sigma_{G} \Gamma^{2}_{G}}{E_{\gamma}[(E^{2}_{\gamma}-E^{2}_{G})^{2}+(E_{\gamma}\Gamma_{G})^{2}]},
  \label{eq:gqer}
\end{equation}
where $\frac{1}{5(\pi\hbar c)^{2}} = 5.20 \times 10^{-8}$~mb$^{-1}$MeV$^{-2}$.  GQER parameters were calculated using the following global systematics: $E_{G} = 63A^{-1/3}$~MeV \cite{speth:81}, $\Gamma_{G} = 6.11 - 0.012A$~MeV \cite{prestwitch:84}, and $\sigma_{G} = 1.5 \times 10^{-4} \cdot \frac{Z^{2}E_{G}^{2}A^{-1/3}}{\Gamma_{G}}$~mb \cite{prestwitch:84}.  Our adopted parameters are listed in Table~\ref{tab:PSF}.

Photon strength corresponding to $M2$ and higher-order multipole transitions were not considered in modeling the neutron-capture $\gamma$-ray cascades in this work due to their expected insignificance.

\section{\label{sec:results}Results and Discussion\protect\\}

Table~\ref{tab:gammas} lists partial $\gamma$-ray production cross sections ($\sigma_{\gamma}$) for 418 $\gamma$ rays associated with 173 levels of the $^{140}$La decay scheme \cite{nica:07} up to an excitation energy of 3009.8~keV.  Of these experimental transitions, 145 are primary $\gamma$ rays originating at the neutron separation energy.  A spin window of $1 \leq J \leq 7$, through both direct and indirect population, is observed.  Using previous information in ENSDF \cite{nica:07}, transitions measured in the prompt $\gamma$-ray spectrum were placed in the $^{140}$La level scheme.  

The nucleus $^{140}$La is odd-odd ($Z=57$, $N=83$) with a relatively high level density.  The $^{139}$La($n,\gamma$) reaction populates many levels based on several possible configurations above the $Z=50$ and $N=82$ shell closures.  The low-lying levels in the residual nucleus are well characterized in terms of quasiparticle couplings between valence protons in the $1g_{7/2}$ and $2d_{5/2}$ orbits with neutrons in the $2f_{7/2}$ and $3p_{3/2}$ orbits \cite{meyer:90,jurney:70}.  Earlier theoretical work suggests these negative-parity states involve the proton-neutron multiplets $|\pi(l_{j_{p}}) \otimes \nu(l_{j_{n}}); |j_{p}-j_{n}|\dots|j_{p}+j_{n}|\rangle$ where $l_{j_{p(n)}}$ and ${j_{p(n)}}$ corresponds to the proton (neutron) orbital and the total angular momentum, respectively, of the aforementioned valence subshells.  There is considerable evidence for mixing between these configurations \cite{kern:67}, leading to a complicated level structure for $^{140}$La.  Indeed, the failure of the Brennan-Bernstein coupling model in odd-odd nuclei \cite{brennan:60} for $^{140}$La, which predicts $6^{-}$ for its ground-state spin-parity rather than $3^{-}$ \cite{nica:07}, is not surprising as the mixing between quasiparticle configurations shifts the level ordering with respect to the pure configuration limits.  Furthermore, theoretical calculations that underestimate configuration mixing fail to accurately reproduce the low-lying spins, e.g., Ref.~\cite{struble:67}.  These observations, together with the underlying theoretical conjecture, are consistent with the observed complicated $\gamma$-ray spectrum shown in Fig.~\ref{fig:LaOspec}.

The $\gamma$-ray transitions observed in this work have been obtained from a singles measurement.  Many of the $\gamma$-ray signals occur in close proximity to one another leading to peaks that are often superimposed on top of one another rendering direct measurement of the intensity difficult.  In these cases, identified by footnotes in Table~\ref{tab:gammas}, branching ratios from ENSDF \cite{nica:07} were used to normalize the cross sections.  The normalization transition for each level, usually the strongest $\gamma$-ray branch, is also indicated.  For certain transitions, also noted in Table~\ref{tab:gammas}, limits on the $\gamma$-ray intensities depopulating a level could only be established through the measured feeding intensity ($\sum_{j=1}^{m} \sigma_{\gamma_{j}}$) to that level.  Here, we assume the depopulation of the level must be at least equal to, or greater than, the observed feeding.  The lower limit for a transition depopulating a given level is obtained as
\begin{equation}
\sigma_{\gamma_{i}} \geq \frac{\sum\limits_{j=1}\limits^{m} \sigma_{\gamma_{j}} (1+\alpha_{j})}{\sum\limits_{i=1}\limits^{n} b_{R_{i}} (1+\alpha_{i})} b_{R_{i}},
\label{eq:aaron_r1}
\end{equation}
where $m$ and $n$ denote the total number of $\gamma$ rays feeding ($j$) and deexciting ($i$) the level, respectively, and $b_{R_{i}}$ is the ENSDF-reported \cite{nica:07} branching ratio.  Limits on $\gamma$-ray cross sections for transitions deexciting the following levels were all determined using Eq.~(\ref{eq:aaron_r1}): 1188.4, 1672.6, 1686.8, 1736.0, 1744.0, 1818.4, 1823.5, 1842.1, 2006.1, and 2125.5~keV.  In addition, for several $\gamma$-ray transitions a contaminant contribution to the observed multiplet, either from the background or another transition of similar energy, was subtracted from the observed peak intensity to arrive at the reported cross sections.  These transitions are also identified with footnotes in Table~\ref{tab:gammas}.

The level energies of $^{140}$La in Table~\ref{tab:gammas} were obtained from a recoil-corrected least-squares fit to the experimental $E_{\gamma}$ data.  These energies compare well with the adopted values in ENSDF \cite{nica:07}.  Spin-parity assignments for the levels involved, $\gamma$-ray transition multipolarities ($XL$) and multipole mixing ratios ($\delta_{\gamma}$) reported in Table~\ref{tab:gammas} were taken from ENSDF \cite{nica:07} where available, and internal-conversion coefficients ($\alpha$) were recalculated with {\small BRICC} \cite{kibedi:08} according to the tabulated transition multipolarities.  The $J^{\pi}$ values could be verified for all but one of the first ten levels up to our established critical energy $E_{c} = 285$~keV (see Sect.~\ref{sec:results.2}).  Unknown transition multipolarities were assumed to be characterized by the lowest multipole order consistent with angular-momentum selection rules cf. Weisskopf single-particle estimates.  Although many of these transitions may, in fact, have mixed-multipole character (with an $E2$ admixture), for most transitions ($E_{\gamma} \gtrsim 200$~keV) this will have a negligible impact on the $\alpha$-corrected cross sections.  On the other hand, for several lower-energy transitions below $E_{c}$, $\delta_{\gamma}$ was adjusted to optimize agreement with the observed $\gamma$-ray intensity balance (see Sect.~\ref{sec:results.4}) and statistical-model calculations.  Similar investigations with $\delta_{\gamma}$ have been carried out previously in $^{181}$W \cite{hurst:15} and $^{186}$Re \cite{matters:16}.

\subsection{\label{sec:results.1}Capture-state $J^{\pi}$ composition\protect\\}

In neutron capture the spin of the capture state ($J_{\rm c.s.}$) in the compound nucleus is determined by $\vec{J}_{\rm c.s.} = \vec{J}_{\rm g.s.} + \vec{l} + \vec{s}$, where $J_{\rm g.s.}$ represents the ground-state spin of the target nucleus, $l$ is the neutron orbital angular momentum and $s = \pm 1/2$ is the neutron spin angular momentum.  The parities of the two states are related by $\pi_{\rm c.s.} = \pi_{\rm g.s.}(-1)^{l}$.  In thermal neutron capture $l=0$ and because the ground state of $^{139}$La is $7/2^{+}$, capture-state resonances in the $^{140}$La compound nucleus have $J^{\pi} = 3^{+}$ and $4^{+}$.  The sum of the cross sections for populating low-spin ($\sigma(-); J^{\pi}=3^{+}$) and high-spin ($\sigma(+); J^{\pi}=4^{+}$) resonances, together with that of any bound resonances ($\sigma(B)$, where $E_{B} < S_{n}$), contribute to the total thermal neutron-capture cross section:
\begin{equation}
\sigma_{0} = \sigma(-) + \sigma(+) + \sigma(B).
\label{eq:aaron_r2}
\end{equation}
In the $^{139}$La target there is a single bound resonance where $E_{B} = -48.63$~eV, $J=4$, and $\sigma(B) = 8.955$~b \cite{mughabghab:06}.  Using the recommended values of $\sigma(-) = 0.084$~b and $\sigma(+) = 0.005$~b for populating the $3^{+}$ and $4^{+}$ capture states above $S_{n}$ \cite{mughabghab:06}, respectively, the spin-fractional composition of the capture state may be deduced from
\begin{equation}
F^{-} = \frac{\sigma(-)}{\sigma_{0}} \quad {\rm and} \quad F^{+} = \frac{\sigma(B) + \sigma(+)}{\sigma_{0}},
\label{eq:aaron_r3}
\end{equation}
where $F^{-} + F^{+} = 1$.  For the statistical-model calculations described in this work we adopt a capture-state composition $J^{\pi}_{\rm c.s.} = 3^{+}(0.9\%) + 4^{+}(99.1\%)$ based on the recommended values above \cite{mughabghab:06}, which agrees well with our experimental data.  

Because the prevailing decay from the capture state proceeds via $E1$ transitions, direct feeding of negative-parity states dominate the decay process and a spin window of $2^{-} \leq J^{\pi} \leq 5^{-}$ may be observed.  However, given that the overwhelming contribution comes from a $4^{+}$ state, this implies the range $3^{-} \leq J^{\pi} \leq 5^{-}$ is favored.  Indeed, Table~\ref{tab:gammas} shows that primary-$\gamma$ decays to final levels below $\sim 1$~MeV with $J^{\pi} = 3^{-}$, $4^{-}$, or $5^{-}$, are, on average, more than an order of magnitude stronger than decays to a $2^{-}$ level.  Using this information together with the observed systematics, we impose the following limitations on $\gamma$-ray multipolarity and final-level $J^{\pi}$ assignments to improve the decay scheme where possible: $J^{\pi} = (3^{-},4^{-},5^{-})$ for levels fed by primary $\gamma$ rays with $\sigma_{\gamma} \geq 0.01$~b assuming $E1$ character and a $4^{+}$ capture state; $J = (2,3,4)$ for levels fed by primary $\gamma$ rays with $\sigma_{\gamma} < 0.01$~b assuming $E1$ or $M1$ character and a $3^{+}$ capture state.  Footnotes are used to identify these assignments in Table~\ref{tab:gammas}.

\subsection{\label{sec:results.2}Constraints on the $^{140}$La decay scheme\protect\\}

\begin{figure*}[t]
  \begin{center}
    \includegraphics[width=\linewidth]{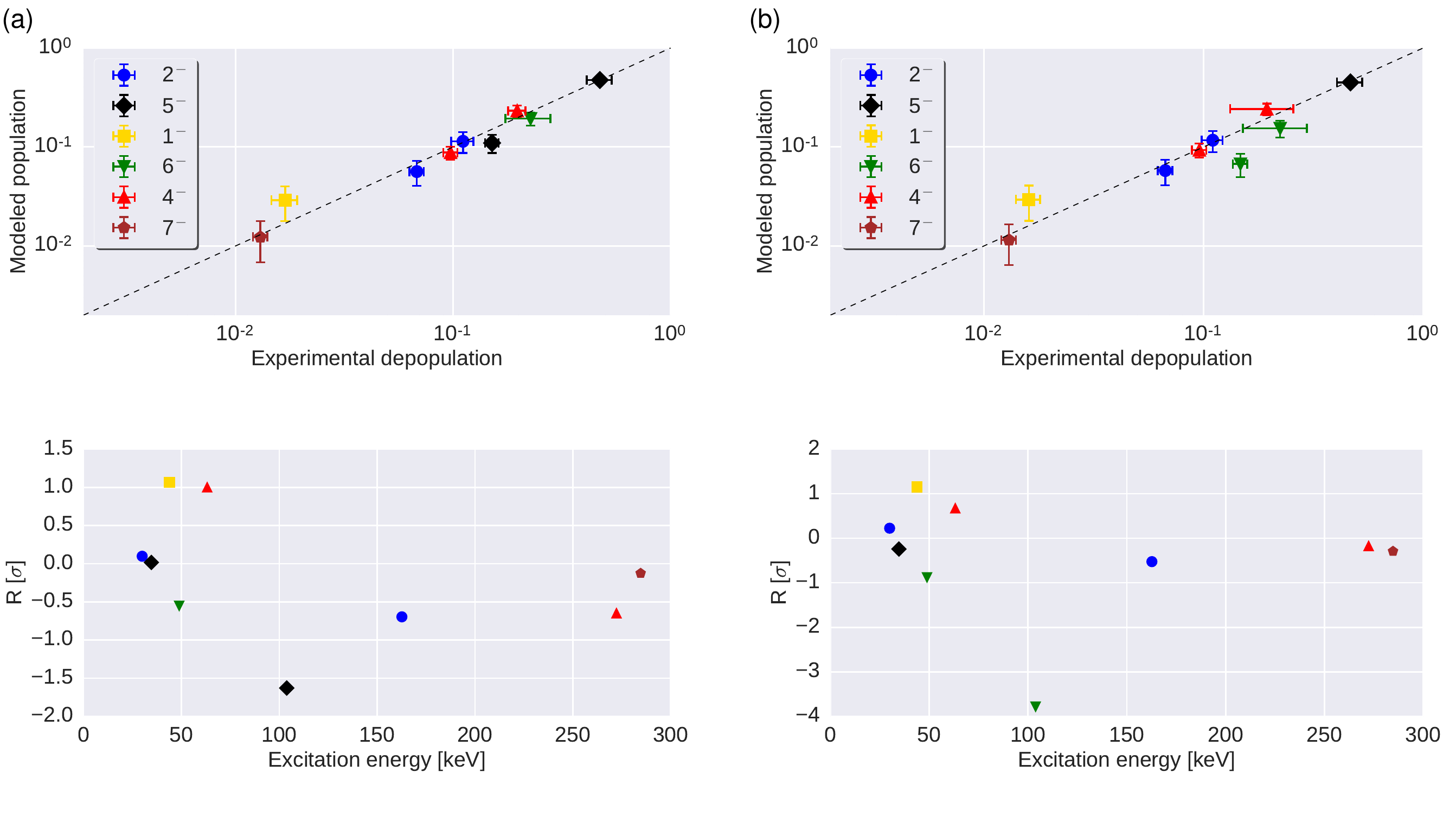}
  \end{center}
  \caption{(Color online)  Upper panels: The simulated population (per neutron capture) of levels below the critical energy of 285~keV versus their experimental depopulation (per neutron capture) assuming using the GLO \cite{kopecky:90} model for the PSF and CTF \cite{ericson:59,gilbert:65} model for the LD using the parametrization of Ref.~\cite{vonegidy:09}.  The $J^{\pi}$ distribution of the levels is indicated on the plot and a dashed-black line of unit slope is drawn for guidance.  Lower panels: Residuals ($R$) between simulations and experimental data as a function of the excitation energy.  The plots correspond to (a) $J^{\pi} = 5^{-}$ and (b) $J^{\pi} = 6^{-}$ assignments for the 103.8-keV level.}
  \label{fig:aaron_pop_depop}
\end{figure*}

In our previous works on light \cite{firestone:13}, medium-mass \cite{firestone:17,choi:14,krticka:08}, and heavy nuclei \cite{hurst:14,hurst:15,matters:16}, it has been demonstrated that comparison of experimental depopulation with simulated population of individual levels is a powerful tool for constraining nuclear structure properties and decay-scheme completeness.  The simulated population of a level $P_{L}^{\rm sim}$ is calculated with {\small DICEBOX} as a fraction per neutron capture, while the measured experimental intensities are determined as absolute cross sections.  Clearly, it is important we compare simulated and experimental quantities with the same units.  To achieve this, we converted the experimental values to intensities per neutron capture of a level $P_{L}^{\rm expt}$ using the relation
\begin{equation}
P_{L}^{\rm expt} = \sum\limits^{n}\limits_{i=1} \frac{\sigma_{\gamma_{i}}(1+\alpha_{i})}{\sigma_{0}},
\label{eq:aaron_r4}
\end{equation}
where $n$ is the number of $\gamma$ rays depopulating a given level.  Preempting the following discussion in this section, a representative comparison of the simulated population with experimental depopulation is shown in Fig.~\ref{fig:aaron_pop_depop}(a) for all levels below $E_{c}$.  This figure was generated using the GLO/CTF model combination for the $E1$ PSF/LD, assuming the PSF parametrization in Table~\ref{tab:PSF} and the LD parametrization of Ref.~\cite{vonegidy:09}.  Porter-Thomas fluctuations from independent nuclear realizations give rise to the uncertainties on the ordinate, while the experimental uncertainty in the measured cross sections generate those on the abscissa.  The uncertainty on $\sigma_{0}$ has not been propagated through because this quantity was only used to normalize $P_{L}^{\rm expt}$.  Agreement between model and experiment is indicated by the close proximity of results to the slope of unit gradient and through their corresponding residuals $R$, defined as $R = P_{L}^{\rm sim} - P_{L}^{\rm expt}$, shown in the lower panel of Fig.~\ref{fig:aaron_pop_depop}.

According to the present ($n,\gamma$) analysis and previous information in ENSDF \cite{nica:07}, we have decided to set $E_{c}$ to 285~keV.  There are 10 levels below this value of $E_{c}$.  The good agreement between simulation and experiment (e.g., Fig.~\ref{fig:aaron_pop_depop}(a)) provide support for choosing a set of adopted models (see Sect.~\ref{sec:results.6}) and the corresponding nuclear structure information for all levels below $E_{c}$: $\gamma$-ray transition energies, internal-conversion coefficients, branching ratios, multipole-mixing ratios, and spin-parity assignments.  

All levels below $E_{c}$ are considered to have firm $J^{\pi}$ assignments, apart from the two ENSDF-reported levels at 92.8 and 106.1~keV \cite{nica:07}, each of which have unknown $J^{\pi}$ assignments and no $\gamma$-decay branches are known.  The only evidence for the 92.8-keV level is from a much earlier $^{139}$La($d,p$) measurement \cite{kern:67} where the authors claim that the observed weakly populated structure may not be real, leading to a tentative assignment.  No other measurements have been able to confirm this level and we find no evidence supporting its assignment in this work.  Furthermore, because all low-lying levels expected from the multiplets of states involving the dominant configurations at low-excitation energy in $^{140}$La have already been accounted for, we support the speculation of Kern \textit{et al}.~\cite{kern:67} and suggest that this level does not exist and have removed it from the decay scheme.  The other level at 106.1~keV was also first tentatively reported in the same $^{139}$La($d,p$) study, again based on a weakly populated structure, and values of $4^{-}, 5^{-}$, or $6^{-}$ are proposed as possible spin-parity assignments for this level \cite{kern:67}.  Level energies consistent with 106.1 and 103.8~keV are also reported in the more recent $^{139}$La($n,2\gamma$) measurement of Vasilieva \textit{et al}. \cite{vasilieva:00}.  However, because of the poor energy resolution associated with that measurement, many of the close-lying intermediate levels could not be unambiguously resolved and these two levels are likely to be an unresolved doublet \cite{nica:07}.  We, therefore, suggest that the weakly populated level at 106~keV in Ref.~\cite{kern:67} is, in fact, the same level measured at 103.8~keV in this work.  

The adopted $J^{\pi}$ assignment for the 103.8-keV level is $6^{-}$ \cite{nica:07}.  Our analysis in this work and previous studies \cite{hurst:14,hurst:15,matters:16,choi:14,krticka:08,firestone:13} shows convincingly that the simulated population of the low-lying levels in many cases depends strongly on their $J^{\pi}$ assignments.  We find a marked reduction in the residual difference between experimental depopulation and simulated population for a $J^{\pi}=5^{-}$ (Fig.~\ref{fig:aaron_pop_depop}(a): $|R| = 1.6\sigma$) over a $J^{\pi}=6^{-}$ (Fig.~\ref{fig:aaron_pop_depop}(b): $|R| = 3.8\sigma$) assignment.  Accordingly, we suggest a revised $5^{-}$ spin-parity hypothesis for this level.  A favorable $E2$ multipolarity for the 554.9-keV $\gamma$ ray feeding this level from the $3^{-}$ level at 658.3~keV accompanies and supports this adjustment, rather than a $\Delta J=3$ transition \cite{nica:07} that previously and otherwise indicates an $M3$ mutlipole.  For all other levels below $E_{c}$, the population-depopulation plot of Fig.~\ref{fig:aaron_pop_depop} does not contradict the proposed $J^{\pi}$ assignments in ENSDF \cite{nica:07}.  

The decay properties for many of the levels presented in Table~\ref{tab:gammas} necessitate further discussion and are described, in turn, below:

\emph{30.0-, 48.9, 63.2-, and 103.8-keV levels.}  All levels are depopulated by low-energy $\gamma$ rays that are difficult to measure in the capture-$\gamma$ spectra, and in some cases below the HPGe threshold.  Partial $\gamma$-ray cross sections for all transitions associated with these levels were, therefore, normalized to the absolute intensity $I_{\gamma}/100n$ data of Meyer \textit{et al}. \cite{meyer:90}, as described earlier in Sect.~\ref{sec:expt.B}.  To help balance the intensities for the 48.9-, 63.2-, and 103.8-keV levels, we deduced optimal internal-conversion coefficients using Eq.~(\ref{eq:aaron_r1}) assuming mixed $M1+E2$ transitions for the 14.24-, 28.53-, and 54.94-keV $\gamma$-ray transitions, respectively.  Accordingly, this optimization process also yields the corresponding mixing ratios for these transitions listed in Table~\ref{tab:gammas}.  An additional 40.6-keV $E2$ $\gamma$-ray transition is also reported to deexcite the 103.8-keV level in Ref.~\cite{meyer:90} with an upper limit of $0.004/100n$.  This transition is not reported in ENSDF \cite{nica:07}.  We are unable to confirm the placement of this transition from the singles capture-$\gamma$ spectra and the plot of Fig.~\ref{fig:aaron_pop_depop}(a) already suggests an approximate equilibrium for the population-depopulation intensity balance.  Nevertheless, we have tentatively listed the 40.6-keV $\gamma$-ray transition with its normalized upper-limit cross section in Table~\ref{tab:gammas} because it does not significantly affect the overall intensity balance of the level (see Sect.~\ref{sec:results.4}).

\emph{34.6-keV level.}  This level is deexcited by a single $E2$ transition direct to the ground state.  Previous absolute intensity measurements for this 34.6-keV $\gamma$ ray have proved problematic, however.  In the work of Jurney \emph{et al}. they report a value of $0.29/100n$, but note that systematic errors arising from $\gamma$-ray self absorption within the source (due to a poorly-defined geometry) may lead to their intensities being off by a factor of two for transitions below 100~keV \cite{jurney:70}.  More recently, Meyer \emph{et al}. report a value of $0.73(2)/100n$ \cite{meyer:90} corresponding to the total intensity of an unresolved region convolved with cerium $K_{\alpha_{1}}$ and $K_{\alpha_{2}}$ $x$-ray doublets observed following the $\beta^{-}$ decay of $^{140}$La to $^{140}$Ce.  Attempts at deconvolving the intensity of the 34.6-keV $\gamma$-ray generate results that differ in magnitude by almost a factor of two and with error bars ranging from $31-48\%$ \cite{meyer:90}.  Because of these ambiguities, we have adopted an intensity balance approach for this level using Eq.~(\ref{eq:aaron_r1}).  Our deduced partial cross section for this transition (Table~\ref{tab:gammas}) yields an average intensity of $0.51(9)/100n$ based on comparison with the data reported in Refs.~\cite{meyer:90,jurney:70}.  This result is significantly less than the total intensity of the region reported in Ref.~\cite{meyer:90}, as should be expected.  It is also consistent with the earlier observation in Ref.~\cite{jurney:70} assuming their correction factor for $\gamma$-ray self absorption.

\emph{43.8-keV level.}  A 13.9-keV $\gamma$ ray is known to deexcite this level.  This transition has previously been reported by Meyer \textit{et al}. with an upper limit of $0.09/100n$ \cite{meyer:90}.  Our reported cross section in Table~\ref{tab:gammas} has been determined using Eq.~(\ref{eq:aaron_r1}) and yields an absolute intensity of $0.04/100n$, consistent with the previous limit.  An additional possible transition at 43.8~keV is reported in Ref.~\cite{meyer:90} with $I_{\gamma} < 0.003/100n$; this transition has also been adopted in ENSDF although its placement is uncertain \cite{nica:07}.  Again, we are unable to provide confirmation of this transition from the capture-$\gamma$ spectra.  However, because we have adopted an intensity-balance argument to determine the total deexcitation of the level, it is possible that deexcitation may occur via an additional decay path and have, therefore, included the normalized cross section of the tentative 43.8-keV $\gamma$ ray as an upper limit in Table~\ref{tab:gammas}.

\emph{318.2 and 322.0-keV levels.}  An intermediate level at 320.2~keV is also reported tentatively in ENSDF with no known $\gamma$-decay branches and without a $J^{\pi}$ assignment \cite{nica:07}.  The only experimental observation of this level comes from a thermal neutron-capture two-$\gamma$ cascade measurement \cite{vasilieva:00}.  However, many of the levels populated in Ref.~\cite{vasilieva:00} are known to be unresolved multiplets \cite{nica:07} and it is speculated that this level is most likely the unresolved multiplet of the nearby 318.2- and 322.0-keV levels, implying its existence is highly questionable.  Furthermore, the observed $\gamma$-ray intensity feeding and deexciting both the 318.2- and 322.0-keV levels is well balanced (see Sect.~\ref{sec:results.4}).  However, in a few instances, a final-level energy of 320.2~keV corresponds to a better excitation-energy fit for some $\gamma$ rays in Table~\ref{tab:gammas}; these cases are identified and may indicate the presence of an unresolved doublet transition to both the 318.2- and 322.0-keV levels.  Thus, we do not include the 320.2-keV level from our measured decay scheme.

\emph{658.3- and 755.3-keV levels.}  Gamma rays with unknown branching ratios at 190.6 and 179.2~keV are reported to deexcite levels at 658.3 and 755.3~keV, respectively \cite{nica:07}.  These transitions were searched for in the prompt spectrum but could not be uniquely identified and are omitted from Table~\ref{tab:gammas}.  It should be noted that the low-energy tail from a strong peak at 181.9~keV from $^{157}$Gd($n,\gamma$) may obscure a $\gamma$-ray line at the reported energy of 179.2(5)~keV.  However, because our measured cross section of 0.500(19)~b is consistent with the expected value of 0.472(38)~b \cite{revay:pgaa}, assuming our measured abundance of $1.03 \times 10^{-3}(7)\%$, this implies any additional transition around this energy is likely to be very weak at best.

\emph{673.0- and 711.7-keV levels.}  A low-energy $\gamma$-ray transition at $\sim 38.7$~keV is reported to deexcite the 711.7-keV level with an unknown branching ratio \cite{nica:07}.  Because this transition is below the HPGe threshold set during the measurement we are unable to determine its partial cross section from the prompt spectra.  The energy region is also partially obscured by an overlapping x~ray.  Due to these limitations we have, therefore, normalized the cross section for the 38.7-keV transition to its absolute intensity reported by Meyer \textit{et al}.~\cite{meyer:90}.  The authors of that work report a weighted average of $0.084(17)/100n$ yielding a normalized value $\sigma_{\gamma} = 0.0058(13)$~b; we have increased the uncertainty on the intensity in Ref.~\cite{meyer:90} to correspond to that of the lowest individual measurement.  The excitation-energy difference $E_{\gamma}=E_{i}-E_{f}$ suggests this transition is most likely to feed the ENSDF-reported level at 673.0-keV \cite{nica:07}.  However, it should be noted that ENSDF also adopts a nearby level at 671.1(5)~keV based on a thermal $^{139}$La($n,2\gamma$) study \cite{vasilieva:00}, hitherto the only measurement of this level.  Although observation of the level at 671.1-keV may be explained due to the poor energy resolution that informs the $^{139}$La($n,2\gamma$) measurement \cite{vasilieva:00}, matters are further complicated given that $\gamma$ rays of distinctly different energies are reported to deexcite these nearly-degenerate levels.  ENSDF lists $\gamma$-ray energies of 565.0(8) and 608.3(6)~keV deexciting the 671.1-keV level, and $\gamma$ rays at 97.1(5) and 638.33(3)~keV from the 673.0-keV level \cite{nica:07}.  We find no evidence for a 97.1-keV transition in our prompt spectra and although we measure an unplaced transition at 637.30(34)~keV, this value is still 1~keV less than the ENSDF value.  Gamma rays of similar energies to those from the 671.1-keV level are, however, observed in the prompt spectra making it difficult to confirm the deexcitation mechanism.  Because the low-energy 38.6948(15)-keV transition from the 711.7-keV level has been accurately determined from a curved-crystal spectrometer measurement \cite{meyer:90} we suggest the final level associated with this decay to be at 673.0~keV.  Unfortunately, we cannot satisfactorily explain the decay of this level and make no attempt to report it$-$or that of a possible level at 671.1~keV$-$in Table~\ref{tab:gammas}.

\emph{744.7-keV level.}  An 86.43(3)-keV $\gamma$-ray is reported to deexcite this level with unknown branching ratio \cite{nica:07}.  Although we measure a nearby peak at 86.93(8)~keV in the prompt spectrum ($\sigma_{\gamma} = 0.0155(15)$~b after correcting for attenuation), a transition of similar energy is also observed in the background spectrum.  Because the intensity balance of the level does not imply significant information is missing and our measured energy is quite far from the ENSDF-reported value, it is difficult to deduce the component, if any, that may be attributed to the deexcitation transition.  Accordingly, we do not list this transition in Table~\ref{tab:gammas}.  

\emph{969.3-keV level.}  ENSDF reports, with uncertain placement, a 968.66(8)-keV transition to the ground state \cite{nica:07}.  In our prompt spectrum the closest candidate transition we find is at 969.27(16)~keV.  We have tentatively included this $\gamma$ ray in Table~\ref{tab:gammas}, however, it should be pointed out that this represents a statistically-significant energy difference of more than $3\sigma$ cf. ENSDF.

\emph{1033.2-keV level.}  The ENSDF-reported $\gamma$ ray at 925.5(15)~keV \cite{nica:07} may populate either the 103.8- or 106.1-keV levels.  Although the 106.1-keV is apparently a better fit, we note that the $\gamma$-ray energy has a large associated uncertainty and this could also allow for a transition to the 103.8-keV level.  Should the 106.1-keV level be fed, we would expect to observe deexcitation of this level.  However, since this is clearly not the case, either in this or previous works \cite{nica:07} and together with our earlier arguments for removal of the 106.1-keV level, we suggest a more likely placement for the poorly-resolved 925.5-keV transition feeds the 103.8-keV level.

\emph{1116.8-, 1264.9-, and 1340.3-keV levels.}  We report improved accuracy regarding $\gamma$-ray energy and branching ratio measurements for all levels cf. ENSDF \cite{nica:07}.  These measurements contribute towards goals outlined in Refs.~\cite{whitepaper:15,zsolnay:12}.  Our measured branching ratios for the 1264.9-keV level are statistically different from those in ENSDF.

\emph{1477.9-keV level.}  ENSDF reports the 1207.1(4)-keV $\gamma$ ray from this level as an undivided doublet.  The same $\gamma$ ray is also reported as the only transition to deexcite the 1479.3-keV level \cite{nica:07}.  The 1477.9-keV level is fed by a primary $\gamma$ ray with $E_{\gamma} = 3683.1$~keV but there is no evidence for primary feeding of the 1479.3-keV level.  This finding is consistent with previous $^{139}$La($n,\gamma$) studies \cite{islam:90}.  We propose that the transition measured at 1208.5-keV is most likely a resolved singlet and, accordingly, suggest revised branching ratios for this level based on $\sigma_{\gamma}$ measurements in Table~\ref{tab:gammas}.  However, the large uncertainty in $\sigma_{\gamma}$ for both the 1208.5- and 1414.7-keV transitions from this level result in branching ratios consistent with those in ENSDF.

\emph{1580.0-keV level.}  A 1420.1(4)-keV $\gamma$-ray is reported in ENSDF as a doublet with undivided intensity; its placement in the level scheme is reported as uncertain.  The same 1420.1-keV $\gamma$-ray is also reported, again with uncertain placement, to deexcite the 1581.5-keV level \cite{nica:07}.  Our analysis reveals primary $\gamma$-ray feeding to the 1580.0-keV level via $E_{\gamma} = 3580.7$~keV, but not to the 1581.5-keV level cf. similar findings in Ref.~\cite{islam:90}.  The ENSDF assignment for the 1581.5-keV level is attributed to the $^{139}$La($n,2\gamma$) measurement \cite{vasilieva:00} and observation of a level at 1583.6(26) keV reported in $^{139}$La($d,p$) \cite{kern:67}.  However, because of the large uncertainties associated with the primary $\gamma$ rays measured in the $^{139}$La($n,2\gamma$) work, many of the intermediate states populated were not unambiguously resolved \cite{vasilieva:00}.  The large uncertainty reported for this level in the $^{139}$La($d,p$) measurement could imply it is an imprecise determination for the 1580.0-keV level reported here; our $J^{\pi}$ assignment for this level is also consistent with a state vector arising from a quasineutron in the $f_{7/2}$ orbital that would be expected for an $l=3$ transfer measured in Ref.~\cite{kern:67}.  For these reasons, we suggest the measured $\gamma$-ray transition at 1420.98(37)~keV is most likely attributable to deexcitation of the 1580.0-keV level alone and have removed the ENSDF-reported level at 1581.5~keV \cite{nica:07} from the decay scheme.  Our measured branching ratios for this level are also statistically different to those in ENSDF.

\emph{1672.6-keV level.}  This level is fed by a 3488.4-keV primary $\gamma$ with $\sigma_{\gamma} = 0.0186(26)$~b.  Previous pair-spectrometry $^{139}$La($n,\gamma$) measurements have also reported this transition \cite{islam:90}.  Because our measurement of the 1621.7-keV deexcitation $\gamma$ ray, the only known decay channel from this level, is part of a doublet centered on $\sim 1623$~keV we impose a lower limit of $\sigma_{\gamma} \geq 0.0186$~b based on the observed primary $\gamma$-ray feeding to the level.  Accordingly, this limit reduces the remaining intensity available for the other 1623.2-keV component of the doublet which deexcites the 1895.7-keV level.

\begin{figure}[t]
  \includegraphics[angle=0,width=0.5\textwidth]{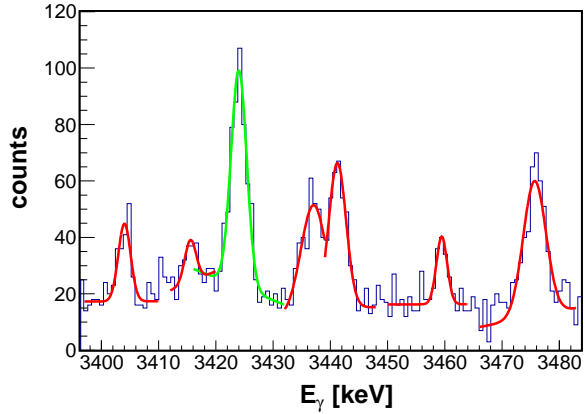}
  \caption{\label{fig:gamma_fit} (Color online) Partial $\gamma$-ray energy spectrum illustrating the fit for the 3424.8-keV primary $\gamma$ ray (green curve).  Representative fits to neighboring $\gamma$ rays are presented for comparison.}
\end{figure}

\emph{1736.0-keV level.}  Earlier work on $^{139}$La($n,\gamma$) reported two nearly-degenerate levels at 1735.6 and 1736.7~keV based on primary $\gamma$ rays observed at 3425.4 and 3424.3~keV, respectively \cite{islam:90}.  Their intensities are reported with a $2.5:1$ ratio in favor of the 3425.4-keV $\gamma$ ray.  In our work, we fit this region with a single transition centered on 3424.8~keV as shown in Fig.~\ref{fig:gamma_fit}.  The full-width half-maximum (FWHM) value for this transition is consistent with those of neighboring $\gamma$ rays, and the residuals between our fitted line shape and measured data also suggest the region is well described by a single transition.  However, upon fitting the region as a doublet, we determine centroids at 3425.4 and 3424.1~keV but with an intensity ratio of 1.35(90).  Furthermore, uncertainties on the FWHM for the doublet transitions are more than a factor of two larger than those of neighboring transitions.  For these reasons we have removed the ENSDF-reported level at 1736.7-keV \cite{nica:07} and allocated the full primary $\gamma$-ray intensity feeding the 1736.0-keV level to that of the transition measured at 3424.8~keV.  As noted in Table~\ref{tab:gammas}, we were only able to obtain limits on the intensities of the transitions deexciting this level using Eq.~(\ref{eq:aaron_r1}).

\emph{1765.7-keV level.}  A single $\gamma$ ray at 1732.2~keV deexcites this level with a measured intensity $\sigma_{\gamma} = 0.0152(37)$~b.  This cross section is consistent with that of the 3395.3-keV primary $\gamma$ ray feeding the level, $\sigma_{\gamma} = 0.0123(30)$~b.  Although a background $\gamma$ ray recorded at 2242.8~keV may give rise to a single-escape peak at $\sim 1732$~keV, it is unlikely to be a source of significant contamination (if any) given the observed balance between feeding and deexcitation of the level.

\emph{1842.1-keV level.}  The 1840.8-keV $\gamma$ ray is reported as an undivided doublet with uncertain placement in ENSDF \cite{nica:07}.  The other doublet deexcites the 1838.9-keV level which is not populated in this measurement.  All $\gamma$ rays from this level are difficult to resolve and intensity limits were determined using Eq.~(\ref{eq:aaron_r1}).  Unfortunately, given the above constraints, we are unable to deduce a unique branching ratio for the 1840.8-keV $\gamma$-ray transition and the limit presented assumes the undivided intensity of ENSDF.

\emph{1859.4-keV level.}  This level is depopulated by a single $\gamma$ ray at 1755.9~keV with a measured cross section of 0.0393(41)~b after resolving for a doublet contribution from the 1818.4-keV level.  However, the deexcitation intensity of this level is (still) very much larger than the feeding intensity of 0.0055(30)~b via the 3301.5-keV primary.  It remains unclear why this discrepancy is so large, although a large contribution from unobserved side feeding could offer one explanation.

\emph{1895.7-keV level.}  A relatively strong 3265.3-keV primary $\gamma$ ray feeds this level ($\sigma_{\gamma} = 0.0515(39)$~b).  The deexcitation of the level is known only to proceed via a single $\gamma$ ray at 1623.2 keV \cite{nica:07}.  In our prompt spectrum we measure a transition centered on 1623.4~keV which has a total intensity of 0.0296(40)~b, nearly a factor of two less than the primary-intensity feeding the level.  However, after removing the intensity attributed to the limit imposed by the 1621.7-keV doublet component that deexcites the 1672.6-keV level, the remaining cross section available for this transition leaves an upper limit of only 0.011~b.  Because it is not physical for the depopulation intensity from a level to be smaller than its population intensity, this may imply there are missing deexcitation $\gamma$ rays associated with this level.  Previous studies in tungsten \cite{hurst:14}, for example, have demonstrated how low-energy $\gamma$ rays ($\lesssim 20$~keV), often difficult to measure experimentally, may explain gaps in the intensity balance of a level.  Indeed, missing (or, hitherto unknown) $\gamma$ rays in general may explain an apparent imbalance.

\emph{1964.1-keV level.}  We report improved $\gamma$-ray energy and branching-ratio measurements for this level.  Our results are statistically consistent with those in ENSDF \cite{nica:07}.

\emph{2044.8-keV level.}  In the adopted literature, a doublet centered on 1993.0~keV  is reported to deexcite this level and the (unobserved) 2041.9-keV level; these placements in the decay scheme are uncertain \cite{nica:07}.  ENSDF gives the undivided intensity for each level.  We could not accurately determine the intensity of a 1993.0-keV transition in the prompt spectrum and have normalized this transition to the stronger well-resolved $\gamma$ ray at 1726.9~keV, assuming the undivided branching ratio from ENSDF of 40(8)\%.

\emph{2069.1-keV level.}  ENSDF reports this level to deexcite via a transition at 1797.5(10)~keV \cite{nica:07}.  In Table~\ref{tab:gammas} we report a more accurate energy at 1795.13(34)~keV (within $3\sigma$).

\emph{2204.6-keV level.}  ENSDF reports consistent branching ratios, albeit with very large error bars, for the 2156.0- and 2164.7-keV $\gamma$ rays that deexcite this level \cite{nica:07}.  Because our cross sections for these transitions also have fairly large error bars, although not as large as those in ENSDF, our branching ratios are also consistent with the ENSDF values.  The 2164.7(18)-keV $\gamma$ ray fits the decay scheme poorly.

\emph{2297.9-keV level.}  Our $\gamma$-ray energy and branching-ratio measurements, listed in Table~\ref{tab:gammas}, for many of the transitions from this level are statistically different ($>1\sigma$) to those in ENSDF \cite{nica:07}.  We do not observe a transition at 2027.2~keV \cite{nica:07} and so have normalized its cross section to that of the best-resolved $\gamma$ line at 2248.3~keV.  All other transitions were extracted directly from the prompt spectrum after ruling out background sources including escape peaks.

\emph{2412.8-keV level.}  ENSDF reports a single deexcitation $\gamma$ ray at $\sim 2094$~keV; its placement in the decay scheme is uncertain \cite{nica:07}.  We measure a weak transition at 2093.8~keV with $\sigma_{\gamma} = 0.0043(33)$~b.  However, the level is fed by a significantly stronger primary at $E_{\gamma} = 2748.1$~keV ($\sigma_{\gamma} = 0.0207(30)$~b).  Because we know this level is populated in $^{139}$La($n,\gamma$) \cite{islam:90}, the lack of deexcitation intensity suggests incomplete decay-scheme information for this level.  Although unlikely to be a significant source of contamination, this scenario is further complicated given that a possible double-escape peak from the 3116.4-keV primary $\gamma$ ray may interfere with the observed intensity in the 2093.8-keV $\gamma$ line.

\subsection{\label{sec:results.3}$J^{\pi}$ assignments for discrete levels above $E_{c}$\protect\\}

In the previous Sect.~\ref{sec:results.2} we described how the statistical model can be used to confirm existing, and infer new, $J^{\pi}$ assignments for discrete levels below the critical energy ($E_{c}$) where we believe the decay scheme is complete.  For levels above $E_{c}$, however, this approach is not possible and we must adopt other nuclear-structure arguments.  Interpretation of the levels above $\sim 700$~keV becomes complicated because of the strong particle-phonon coupling expected in $^{140}$La where a linear combination of an impracticably large number of spherical shell-model basis vectors is required to describe the overall state vector \cite{kern:67}.  The lower-lying levels, however, are better understood in terms of configurations based on couplings involving $1g_{7/2}$ and $2d_{5/2}$ quasiprotons with $2f_{7/2}$ and, to a lesser extent, $3p_{3/2}$ quasineutrons.  The fact that all measured angular distributions to levels below 600~keV are consistent with an $l=3$ transfer demonstrates the dominance of the $\nu(2f_{7/2})$ orbital at low excitation energy \cite{kern:67}.  Above this energy, $l=1$ transfers have been observed for several levels \cite{kern:67} indicating the onset of the $\nu(3p_{3/2})$ orbital; this is also to be expected according to systematics of the neighboring isotonic odd-$A$ nuclei where the $\nu(3p_{3/2})$ orbital begins to appear around 700~keV, e.g., $^{141}$Ce \cite{fulmer:62}.  Using this information, we can now assign level properties to states below 600~keV in terms of $J^{\pi}$ assignments from the expected multiplets of levels not used to describe other levels.

\emph{322.0-keV level.}  This level is currently reported with $J^{\pi} = (5^{-}, 6^{-})$ in ENSDF \cite{nica:07}.  Fourteen low-lying levels are expected from the two configurations based on couplings with the $2f_{7/2}$ quasineutron:
\begin{eqnarray}
\nonumber
&|&\pi(1g_{7/2}) \otimes \nu(2f_{7/2}); J^{\pi} = 0^{-},\dots,7^{-}\rangle\\
\nonumber
&|&\pi(2d_{5/2}) \otimes \nu(2f_{7/2}); J^{\pi} = 1^{-},\dots,6^{-}\rangle
\end{eqnarray}
Because we are suggesting a $J^{\pi}=5^{-}$ assignment for the 103.8-keV level and the other $5^{-}$ assignment is already exhausted through the 34.6-keV level, this leaves a $6^{-}$ assignment as the only remaining possibility from the above octuplet and sextuplet of levels, with the other $6^{-}$ assignment occupied by the 48.9-keV level.  Particle-phonon coupling is demonstrably unimportant at such low excitation energies in $^{140}$La \cite{jurney:70}, and the $\nu(3p_{3/2})$ orbital is unlikely to play a significant role through four-quasiparticle admixtures in this regime.  Furthermore, a $6^{-}$ level cannot be generated by coupling quasiparticles in the $\nu(3p_{3/2})$ orbital with those in the available proton orbitals.  We can, thus, characterize the state vector for the $6^{-}$ level assuming the mixed configuration: $\alpha^{2}|\pi(1g_{7/2}) \otimes \nu(2f_{7/2}); J^{\pi} = 6^{-}\rangle \oplus \beta^{2}|\pi(2d_{5/2}) \otimes \nu(2f_{7/2}); J^{\pi} = 6^{-}\rangle$, where the mixing amplitudes $\alpha^{2}+\beta^{2} = 1$.  Direct reaction theory shows that the cross section for a state produced in a ($d,p$) reaction is proportional to its spectroscopic factor ($S$), thus, $S \propto \alpha^{2}$ for $\pi(1g_{7/2})$ occupancy, and $S \propto \beta^{2}$ for $\pi(2d_{5/2})$ occupancy.  In the $^{139}$La($d,p$) work the authors measure $|\alpha|=1.0$ for the 48.9-keV $6^{-}$ level \cite{kern:67} and do not observe a second $6^{-}$ level.  However, neutron capture, which is generally expected to proceed through the formation of a compound nucleus is not subject to this selectivity.  We propose the observed 322.0-keV level is the final level of the sextuplet to be accounted for with $J^{\pi}=6^{-}$ and corresponds to an essentially pure $\pi(2d_{5/2})$ quasiproton configuration.

For higher-lying levels, we must use additional information to pin down possible $J^{\pi}$ assignments.  From systematics of the observed $\gamma$-ray strengths discussed in Sect.~\ref{sec:results.1}, we may impose ranges of permissible $J^{\pi}$ assignments to levels fed by primary $\gamma$ rays.  And because we expect dipole and quadrupole transitions to dominate the secondary component of the $\gamma$-decay scheme \cite{jurney:70,bogdanovic:88}, we may then use decay properties of the associated intermediate and final levels to further constrain the range of $J^{\pi}$ values or even deduce unique solutions in some cases.  Assignments deduced using these methodologies are flagged for several levels in Table~\ref{tab:gammas} and a summary of the important findings is discussed below.

New $J^{\pi}$ assignments are proposed for 123 levels based on the observed primary $\gamma$-ray feeding; 93 of these assignments are deduced assuming a likely $E1$ transition from a $4^{+}$ capture state and the remaining 30 assignments have been inferred assuming either an $E1$ or $M1$ transition from a $3^{+}$ capture state.  Of the 93 levels fed from the $4^{+}$ capture state, 23 of the $J^{\pi}$ assignments are constrained further according to angular-momentum selection rules with respect to the observed decay modes and final states associated with the level populated by the primary $\gamma$ ray.  Similarly, we were able to further constrain the $J^{\pi}$ assignment for five of the levels populated from a $3^{+}$ capture state.  Based on this analysis we report unique assignments for six levels: 796.3 ($2^{-}$), 1547.9 ($4^{-}$), 1652.5 ($4^{-}$), 2006.1 ($3^{-}$), 2018.2 ($4^{-}$), and 2120.6~keV ($4^{-}$).  The assignments for the 1652.5- and 2006.1-keV levels are tentative, however, because they are deexcited by transitions with uncertain placement in the decay scheme \cite{nica:07}, while that for the 1547.67-keV level is tentative owing to the large uncertainty on both the energy and intensity of its associated primary $\gamma$ ray.  The $J^{\pi}$ assignment for the 796.3-keV level is also tentative (see below).

From the analysis of the secondary $\gamma$ rays alone deexciting levels which are not connected to a primary $\gamma$ ray, we have determined new $J^{\pi}$ assignments for an additional eight levels.  For the 1486.0- and 1496.3-keV levels, we deduce $J^{\pi}=4^{-}$ in each case.  This unique assignment is the only plausible candidate under the restriction that observed transitions are $E1$, $M1$, or $E2$ \cite{jurney:70}, because each level deexcites to a $2^{-}$ and $6^{-}$ level requiring the accommodation of stretched quadrupoles.  We have also confirmed adopted $J^{\pi}$ assignments (or ranges) for 23 levels above $E_{c}$ on the basis of observed feeding and/or deexcitation of the level involved, together with known nuclear structure properties of all initial and final levels associated with the measured transitions.  The confirmed assignments are indicated in Table~\ref{tab:gammas}.

Further discussion is warranted for $J^{\pi}$ assignments concerning certain other levels in Table~\ref{tab:gammas}:

\textit{777.4-, 1550.9-, 1700.6-, 1736.0-, and 1842.1-keV levels.}  All levels are reported with firm $4^{-}$ assignments in the adopted literature \cite{nica:07}.  These unique assignments are consistent with the currently adopted $6^{-}$ assignment for the 103.8-keV level that is populated in the decay sequence of all levels since $\Delta J \leq 2$ for all $\gamma$-ray transitions.  However, because we have established a $5^{-}$ assignment for the 103.8-keV level according to our statistical-model calculations, this also introduces $3^{-}$ as a possible alternative assignment consistent with $\Delta J \leq 2$ for all $\gamma$-ray transitions.  Thus, for these levels we propose $J^{\pi} = (3^{-},4^{-})$.

\textit{796.3-keV level.}  A $(2^{-})$ assignment is proposed in Ref.~\cite{jurney:70} on the basis of deexcitation $\gamma$ rays feeding $0^{-}$, $2^{-}$, and $3^{-}$ levels.  However, the authors acknowledge a lack of supporting evidence from a coincidence measurement and the transition to the $0^{-}$ level has been reported in ENSDF with uncertain placement \cite{nica:07}.  Because we do not measure coincidence data in this work we are unable to affirm this claim.  The ENSDF-reported branching ratio 61(15)\% \cite{nica:07} for the 215-keV transition to the $0^{-}$ level at 581~keV implies $\langle \sigma_{\gamma} \rangle = 0.0264(68)$~b.  Our capture-$\gamma$ spectrum reveals no evidence for a transition at this energy unless it is obscured by the low-energy tail of the much stronger 218.2-keV $\gamma$ ray from the 322.0-keV level, and so we have declined to include this transition in Table~\ref{tab:gammas}.  Furthermore, it is not possible to confirm observation of the 581-keV level (only expected to be weakly populated at best) because the strongest $\gamma$ branch at 537.3~keV, observed in $\beta^{-}$-decay studies \cite{meyer:90}, cannot be clearly identified due to the proximity of stronger peaks overlapping in this region, while all other branches from the 581-keV level are considerably weaker \cite{nica:07}.  Earlier $^{139}$La($n,\gamma$) measurements \cite{firestone:06,islam:90,jurney:70} also failed to populate the pure $\pi(1g_{7/2})$ 581-keV level, known only from $\beta^{-}$ decay \cite{meyer:90} and $^{139}$La($d,p$) \cite{kern:67}.  Although we are unable to improve upon the tentative assignment for the 796.3-keV level, a $(2^{-})$ value seems appropriate because the level is fed by a weak primary $\gamma$ ray consistent with a $3^{+}$ capture state and comparable in strength to the primaries feeding the other low-lying $2^{-}$ levels at 30.0 and 162.7~keV.  

\textit{912.2-keV level.}  Seven $\gamma$ rays are reported to deexcite this level \cite{nica:07} for which we observe no primary feeding.  In Ref.~\cite{jurney:70} the authors report that observation of the transition to both the $1^{-}$ and $6^{-}$ levels must be an accidental energy fit.  We measure a strong 868.1-keV transition that fits the $1^{-}$ level and report a normalized cross section for the 863.3-keV transition that fits the $6^{-}$ level.  We concur with the findings of Ref.~\cite{jurney:70} and note that exclusion of the transition to the $1^{-}$ level implies $J^{\pi} = 4^{-}$, whereas if the transition to the $6^{-}$ level is disallowed then we may accommodate all other transitions assuming $J^{\pi} = 2^{-}$ or $3^{-}$.  Because we have insufficient information to rule out either transition we tentatively assign $J^{\pi} = (2^{-}, 3^{-}, 4^{-})$ for this level.

\textit{914.1-, 1683.9-, and 1823.5-keV levels.}  These levels all decay via the \textit{a priori} uniquely determined $J^{\pi} = 6^{-}$ level at 322.0-keV (cf. $J^{\pi}=(5^{-},6^{-})$ in ENSDF \cite{nica:07}).  This assignment further constrains the currently adopted ranges for these levels from $J^{\pi}=(3^{-},4,5^{-})$ \cite{nica:07} to a narrower window $J^{\pi} = (4^{-},5^{-})$ owing to the exclusion of $\Delta J=3$ $\gamma$-ray transitions.

\textit{917.8-keV level.}  ENSDF reports a relatively weak 341.8-keV transition to a level at 575.9 keV with $J^{\pi} = (2^{-},3^{-})$ \cite{nica:07}.  This transition is attributed to an earlier thermal neutron-capture measurement \cite{firestone:06}, however, there is no evidence for this transition in our spectrum.  A single 413.2-keV $\gamma$ ray is also reported to deexcite the 575.9-keV level in Ref.~\cite{firestone:06}.  The closest transition we observe in our spectrum occurs at 414.9~keV; these two results are at odds by more than $3\sigma$.  No other measurement reports a level at 575.9~keV and our study of the singles capture-$\gamma$ data is inconclusive in ascertaining its experimental validity.  From a theoretical perspective, a two-quasiparticle model demonstrably describes the state vectors for the first 14 levels through to the 581-keV level \cite{jurney:70}, and there is no indication of the onset of the $3p_{3/2}(\nu)$ orbital until above 600~keV \cite{kern:67}, thus, suggesting negligible phonon or four-quasiparticle admixture at low excitation.  Accordingly, it is difficult to explain the existence of an additional (and unexpected) level below the pure $1g_{7/2}(\pi)$ $0^{-}$ level and we do not include transitions to or from the 575.9-keV level in our decay scheme.

\textit{941.7-, 969.3-, 1264.9-, 1442.6-, and 1449.0-keV levels.}  None of these levels are populated by primary $\gamma$ rays.  However, assuming they decay solely via $E1$, $M1$, or $E2$ transitions, this suggests a broader range of $J^{\pi}$ values is possible compared to the reported range in ENSDF \cite{nica:07}.  The 941.7-keV level is reported without assignment and although values of $J^{\pi} = (1^{-},2,3,4,5^{-})$ are consistent with the current decay-mode assumption, the large number of permutations do not allow us to constrain the result in a meaningful manner.  Similar findings are manifest for the other levels.  For these levels, we list the adopted ENSDF $J^{\pi}$ assignments in Table~\ref{tab:gammas} while noting that a broader range of values is permissible according to the current $\gamma$-decay analysis.

\textit{1879.4-, 1971.5-keV levels.}  For similar reasons outlined earlier, the reassignment of the 103.8-keV level to $J^{\pi}=5^{-}$ extends the range of permutations for the 1879.4-keV level to $J^{\pi} = (3^{-},4^{-},5^{-})$ (cf. $J^{\pi} = (4^{-},5^{-})$ in ENSDF \cite{nica:07}).  Also, because we have deduced $J^{\pi} = 6^{-}$ for the 322.0-keV level and the highest expected multipole-order decay is $E2$, this implies a range $J^{\pi} = (4^{-},5^{-})$ for the 1971.5-keV level (cf. $J^{\pi} = (4^{-},5)$ in ENSDF \cite{nica:07}).  In addition, the observed relatively strong $E1$-primary feeding to each level mandates a negative parity for both levels.


\clearpage
\newpage



\begin{minipage}{2.0\linewidth}

\renewcommand{\footnoterule}{}
\footnotetext[1]{$E_{\gamma}$ taken from adopted value in ENSDF \cite{nica:07}.}
\footnotetext[2]{Cross section deduced from intensity balance and ENSDF branching ratios \cite{nica:07}.}
\footnotetext[3]{Multiplet resolved using ENSDF branching ratios \cite{nica:07}.}
\footnotetext[4]{$\gamma$-ray transition used for level branching-ratio normalization.}
\footnotetext[5]{Contaminant contribution subtracted from multiplet.}
\footnotetext[6]{Multiplet transition; limit estimated from observed $\gamma$ intensity feeding level according to Eq.~(\ref{eq:aaron_r1}).}
\footnotetext[7]{Newly proposed $J^{\pi}$ assignment based on measured primary $\gamma$-ray feeding the level with likely $E1$ character assuming an initial $4^{+}$ capture state.}
\footnotetext[8]{$E_{\gamma}$ multiply placed in ENSDF; undivided intensity is given.}
\footnotetext[9]{Transition placement in ENSDF decay scheme is uncertain \cite{nica:07}.}
\footnotetext[10]{Multiply-placed transition in ENSDF decay scheme; undivided branching ratios \cite{nica:07} used to normalize $\sigma_{\gamma}$.}
\footnotetext[11]{Level cited as a possible multiplet in ENSDF \cite{nica:07}.}
\footnotetext[12]{Level cited with tentative placement in ENSDF decay scheme \cite{nica:07}.}
\footnotetext[13]{$\alpha$ deduced from ENSDF-reported $\delta_{\gamma}$ \cite{nica:07}.}
\footnotetext[14]{$\alpha$ taken from ENSDF \cite{nica:07}.}
\footnotetext[15]{$\delta_{\gamma}$ deduced from ENSDF-reported $\alpha$ \cite{nica:07}.}
\footnotetext[16]{Cross section corrected for $^{157}$Gd($n,\gamma$) contribution to observed peak.}
\footnotetext[17]{Newly proposed $J^{\pi}$ assignment, or range, based on decay modes to final states.}
\footnotetext[18]{Cross section normalized by comparison with absolute intensity per 100 neutron captures measured in Ref.~\cite{meyer:90}}
\footnotetext[19]{Newly proposed $J$ assignment based on observed weak primary $\gamma$-ray feeding the level with likely $M1$ or $E1$ character assuming an initial $3^{+}$ capture state.}
\footnotetext[20]{$E_{\gamma}$ taken from Ref.~\cite{meyer:90}.}
\footnotetext[21]{$\alpha, \delta_{\gamma}$ deduced from $\gamma$-ray intensity balance.}
\footnotetext[22]{Adopted $J^{\pi}$ assignment \cite{nica:07}; $\gamma$-decay analysis indicates a broader range of possible values (see text).}
\footnotetext[23]{$J^{\pi}$ assignment confirmed based on decay properties of associated transitions feeding and/or deexciting level (and corresponding levels involved).}
\footnotetext[24]{Previously reported $E_{f} = 106.1$~keV \cite{nica:07}; level not observed in this work.}
\footnotetext[25]{Previously reported $E_{f} = 320.2$~keV \cite{nica:07}; level not observed in this work.}
\footnotetext[26]{$J^{\pi}$ assignment confirmed based on statistical-model calculations.}
\footnotetext[27]{Newly proposed $J^{\pi}$ assignment based on statistical-model calculations.}
\footnotetext[28]{Newly proposed $J^{\pi}$ assignment based on mixed configuration: $|\pi(1g_{7/2}) \otimes \nu(2f_{7/2}); J^{\pi}=6^{-} \rangle$ and $|\pi(2d_{5/2}) \otimes \nu(2f_{7/2}); J^{\pi}=6^{-} \rangle$.}
\footnotetext[29]{One of these transitions is an accidental energy fit to the level \cite{jurney:70}.}

\end{minipage}

\clearpage
\newpage


\subsection{\label{sec:results.5}Neutron-separation energy for $^{140}$La\protect\\}

The primary $\gamma$-ray transitions listed in Table~\ref{tab:gammas} were used to determine the neutron-separation energy $S_{n}$ for $^{140}$La.  An expanded region of the spectrum corresponding to these primaries is shown in Fig.~\ref{fig:LaOspec}.  The value of $S_{n}$ may be determined from the measured primaries according to the final level ($E_{f}$) populated by the $\gamma$ ray:
\begin{equation}
S_{n} = E_{\gamma} + E_{f} + E_{r},
\label{eq:sn1}
\end{equation}
where $E_{r} = E_{\gamma}^{2}/2Mc^{2}$ accounts for the recoil energy of the compound nucleus of mass $M$.  A weighted least-squares fit of the recoil-corrected $\gamma$-ray energies yields $S_{n} = 5161.005(21)$~keV for the capture state in $^{140}$La.  This result is consistent with the adopted value of $5160.98(4)$~keV from the recent atomic mass evaluation by Wang \textit{et al}. \cite{wang:17}, although our uncertainty represents a factor of two improvement in precision.

\subsection{\label{sec:results.4}Intensity balance\protect\\}

\begin{figure}[t]
  \includegraphics[width=1.0\linewidth]{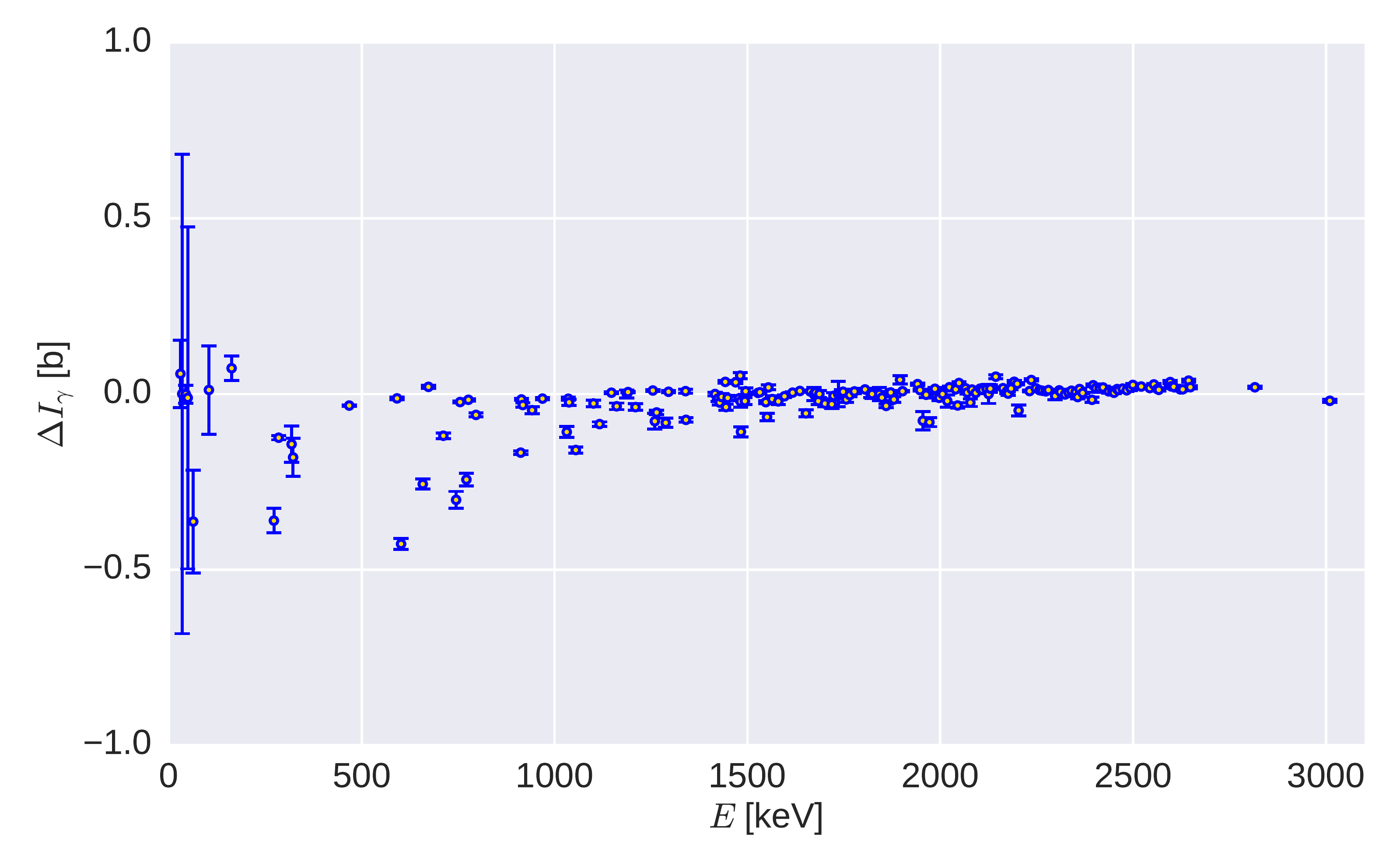}
  \caption{\label{fig:intensity} (Color online) Plot of the total $\gamma$-ray intensity balance ($\Delta I_{\gamma} = \sum I_{\gamma}(\rm in) - \sum I_{\gamma}(\rm out)$) for all levels observed in $^{140}$La.}
\end{figure}

A $\gamma$-ray intensity balance is determined for all levels observed in this work according to
\begin{equation}
\Delta I_{\gamma} = \sum\limits_{i=1}^{m} I_{\gamma_{i}} ({\rm in}) - \sum\limits_{j=1}^{n} I_{\gamma_{j}} ({\rm out}),
\label{eq:intensity}
\end{equation}
where $I_{\gamma} = \sigma_{\gamma}(1+\alpha)$ and represents the total $\gamma$-ray intensity corrected for internal conversion.  Here, $m$ and $n$ denote the observed number of $\gamma$ rays populating ($i$) and depopulating ($j$) a given level, respectively.  Using Eq.~(\ref{eq:intensity}) together with data from Table~\ref{tab:gammas}, we have determined the intensity balance for each level and the results are tabulated in Table~\ref{tab:intensity} and plotted in Fig.~\ref{fig:intensity}.  These results generally show that $\sum I_{\gamma}({\rm out}) \gtrsim \sum I_{\gamma} ({\rm in})$ for most levels below $E_{L} \lesssim 1500$~keV.  For these levels, it is likely that we observe all (or at least the vast majority) of the decaying intensity.  However, because there is unobserved side feeding to some of these levels, we do not always account for the full amount of feeding intensity and in these circumstances $\Delta I_{\gamma} < 0$.  Above this energy, the situation is reversed in several instances and $\Delta I_{\gamma} > 0$.  This behavior is understandable because the feeding of certain levels is often dominated by primary $\gamma$-ray transitions but only a fraction of the decays are observed, and in many cases no decaying intensity is observed.  Table~\ref{tab:intensity} indicates a total of 71 levels (not including the ground state) populated by primary $\gamma$ rays with no known deexcitation $\gamma$ rays, i.e., $\sum I_{\gamma}({\rm out}) = 0$.  

For certain levels in Tables~\ref{tab:gammas} and \ref{tab:intensity} it was necessary to balance the missing intensity $\Delta I_{\gamma}$ from $\gamma$ rays that were obscured by neighboring doublets, or in a few cases, adjusting $\delta_{\gamma}$ for mixed $M1 + E2$ transitions (within the limits of pure multipoles) where the adopted values \cite{nica:07} could not recover sufficient intensity to account for imbalance from the observed feeding to the levels.  Specific cases are highlighted and discussed earlier in Sect.~\ref{sec:results.2}.

Figure~\ref{fig:intensity} and the data from Table~\ref{tab:intensity} show the 162.7-keV level to be slightly over populated, although $\Delta I_{\gamma} = 0$ at $2\sigma$.  The largest decay contribution from this level comes from the mixed $M1+E2$ transition direct to the ground state.  Increasing the mixing ratio closer to the pure $E2$ limit will favor an improved overall intensity balance for the level.  Also, if the $M1$ transitions currently reported to deexcite this level \cite{nica:07} were to have a mixed $E2$ component, this would also improve the intensity balance.  Although the present discrepancy is not very large, our current results suggest internal conversion may have been underestimated for some of the deexcitation transitions from this level.

The total observed intensity populating the ground state is 8.58(50)~b, in agreement with the adopted total radiative thermal neutron-capture cross section, $\sigma_{0} = 9.04(4)$~b \cite{mughabghab:06}.  This implies that the dominant transitions feeding the ground state have been experimentally verified.  However, the total observed intensity of primary $\gamma$ rays depopulating the capture state ($\sum \sigma_{\gamma}^{\rm p}$) at $S_{n}$ is only 5.346(62)~b.  By definition $\sum \sigma_{\gamma}^{\rm p} = \sigma_{0}$, meaning only $\sim 60\%$ of the total decay intensity out of the capture state is accounted for compared to the observed intensity feeding the ground state.  This implies a significant fraction of primary $\gamma$ rays are weak low-energy transitions to levels in the quasicontinuum.

\AltTable
\begin{longtable}{p{2.0cm}p{2.0cm}p{2.0cm}p{2.0cm}}
  \caption{\label{tab:intensity} Level energies and intensity balance obtained using Eq.~(\ref{eq:intensity}) for all levels in $^{140}$La.  All $I_{\gamma}$ values are corrected for internal conversion, i.e., $I_{\gamma} = \sigma_{\gamma}(1+\alpha)$.}\\
  \toprule
  $E$ (keV) & $\sum I_{\gamma}$(in) [b] & $\sum I_{\gamma}$(out) [b] & $\Delta I_{\gamma}$ [b] \\ \hline\endfirsthead
  \caption{(\textit{Continued}.)}\\ \hline\hline
  $E$ (keV) & $\sum I_{\gamma}$(in) [b] & $\sum I_{\gamma}$(out) [b] & $\Delta I_{\gamma}$ [b] \\ \hline\endhead
  \hline\endfoot
  \hline\hline\endlastfoot

    0.00         & 8.58(50)     & 0            & $8.58(50)    $ \\
    29.9642(6)        & 1.104(35)    & 1.047(89)    & $0.058(96)   $ \\
    34.6464(9)        & 4.46(48)     & 4.46(48)     & $0.00(68)    $\footnotemark[1] \\
    43.85(3)        & 0.150(18)    & 0.150(18)    & $0.000(25)   $\footnotemark[1] \\
    48.8848(20)        & 2.13(12)     & 2.14(47)     & $-0.01(49)   $ \\
    63.1791(8)        & 1.495(46)    & 1.86(14)     & $-0.36(15)   $ \\
    103.8291(20)       & 1.416(51)    & 1.40(12)     & $0.01(13)    $ \\
    162.657(11)       & 0.713(26)    & 0.640(23)    & $0.073(35)   $ \\
    272.3071(21)       & 0.554(25)    & 0.914(24)    & $-0.360(35)  $ \\
    284.657(10)       & 0            & 0.1244(58)   & $-0.1244(58) $ \\
    318.220(7)       & 1.000(41)    & 1.143(32)    & $-0.143(52)  $ \\
    322.047(11)       & 0.786(42)    & 0.967(35)    & $-0.180(54)  $ \\
    467.65(4)       & 0.00172(58)  & 0.0346(17)   & $-0.0329(18) $ \\
    591.44(10)       & 0            & 0.0124(36)   & $-0.0124(36) $ \\
    602.036(11)       & 0.1485(60)   & 0.575(15)    & $-0.427(16)  $ \\
    658.283(12)       & 0.1564(94)   & 0.413(11)    & $-0.256(14)  $ \\
    672.985(22)       & 0.0204(46)   & 0            & $0.0204(46)  $ \\
    711.680(22)       & 0.0063(15)   & 0.1251(77)   & $-0.1188(79) $ \\
    744.708(17)       & 0.257(12)    & 0.558(21)    & $-0.301(24)  $ \\
    755.29(15)       & 0            & 0.0228(26)   & $-0.0228(26) $ \\
    771.431(12)       & 0.282(13)    & 0.525(12)    & $-0.244(18)  $ \\
    777.38(13)       & 0            & 0.0165(33)   & $-0.0165(33) $ \\
    796.27(3)       & 0.0150(29)   & 0.0744(44)   & $-0.0595(53) $ \\
    912.159(18)       & 0            & 0.1669(49)   & $-0.1669(49) $ \\
    914.08(14)       & 0            & 0.0155(26)   & $-0.0155(26) $ \\
    917.78(6)       & 0            & 0.0313(58)   & $-0.0313(58) $ \\
    941.73(10)       & 0            & 0.046(10)    & $-0.046(10)  $ \\
    969.27(16)       & 0            & 0.0130(27)   & $-0.0130(27) $ \\
    1033.20(23)      & 0            & 0.108(16)    & $-0.108(16)  $ \\
    1035.63(3)      & 0.0138(23)   & 0.0273(41)   & $-0.0135(47) $ \\
    1038.71(11)      & 0            & 0.0238(84)   & $-0.0238(84) $ \\
    1055.045(9)      & 0.0236(31)   & 0.1830(80)   & $-0.1594(86) $ \\
    1101.06(8)      & 0.0322(32)   & 0.0592(86)   & $-0.0270(92) $ \\
    1116.76(5)      & 0.0321(31)   & 0.1178(56)   & $-0.0856(64) $ \\
    1147.0(9)      & 0.0037(20)   & 0            & $0.0037(20)  $ \\
    1162.7(3)      & 0.00159(60)  & 0.0366(93)   & $-0.0350(93) $ \\
    1188.4(4)      & 0.0150(40)   & 0.015(11)    & $0.000(12)   $ \\
    1190.6(7)       & 0.0055(21)   & 0            & $0.0055(21)  $ \\
    1210.1(4)      & 0.0249(34)   & 0.0622(87)   & $-0.0373(94) $ \\
    1254.2(4)      & 0.0100(22)   & 0            & $0.0100(22)  $ \\
    1260.15(5)      & 0.0524(37)   & 0.130(22)    & $-0.077(23)  $ \\
    1264.91(12)      & 0            & 0.0525(62)   & $-0.0525(62) $ \\
    1286.52(10)      & 0.0063(21)   & 0.088(13)    & $-0.082(13)  $ \\
    1294.82(10)      & 0.0064(30)   & 0            & $0.0064(30)  $ \\
    1339.78(24)      & 0.0191(25)   & 0.0109(48)   & $0.0082(54)  $ \\
    1340.33(8)      & 0            & 0.0735(61)   & $-0.0735(61) $ \\
    1416.31(24)      & 0.0222(35)   & 0.0222(35)   & $0.0000(50)  $ \\
    1423.16(17)      & 0.0402(43)   & 0.0531(70)   & $-0.0129(82) $ \\
    1426.01(14)      & 0.0161(38)   & 0.0403(54)   & $-0.0242(66) $ \\
    1433.63(7)      & 0.0686(64)   & 0.0769(65)   & $-0.0083(91) $ \\
    1442.61(10)      & 0            & 0.0373(89)   & $-0.0373(89) $ \\
    1443.04(20)      & 0.0345(40)   & 0            & $0.0345(40)  $ \\
    1449.0(12)      & 0            & 0.0111(71)   & $-0.0111(71) $ \\
    1469.91(17)       & 0.0333(39)   & 0            & $0.0333(39)  $ \\
    1477.85(18)      & 0.0061(26)   & 0.0191(59)   & $-0.0130(64) $ \\
    1481.46(4)      & 0.1336(73)   & 0.0813(53)   & $0.0523(90)  $ \\
    1482.6(8)      & 0            & 0.021(16)    & $-0.021(16)  $ \\
    1486.0(11)      & 0            & 0.108(15)    & $-0.108(15)  $ \\
    1495.49(4)      & 0.1283(71)   & 0.1204(58)   & $0.0079(92)  $ \\
    1496.3(10)      & 0            & 0.020(10)    & $-0.020(10)  $ \\
    1527.8(9)      & 0.0032(19)   & 0            & $0.0032(19)  $ \\
    1532.3(6)      & 0.0049(20)   & 0            & $0.0049(20)  $ \\
    1547.91(16)      & 0.0028(24)   & 0.0259(42)   & $-0.0231(48) $ \\
    1550.91(9)      & 0.0485(50)   & 0.1140(97)   & $-0.065(11)  $ \\
    1554.81(10)      & 0.0560(54)   & 0.0372(43)   & $0.0188(69)  $ \\
    1564.41(22)      & 0.0173(27)   & 0.0315(83)   & $-0.0142(88) $ \\
    1579.99(24)      & 0.0123(26)   & 0.0333(80)   & $-0.0210(84) $ \\
    1597.1(3)      & 0.0119(22)   & 0.0183(27)   & $-0.0064(35) $ \\
    1617.1(7)      & 0.0039(20)   & 0            & $0.0039(20)  $ \\
    1636.6(6)      & 0.0086(23)   & 0            & $0.0086(23)  $ \\
    1652.5(7)      & 0.0055(24)   & 0.0606(94)   & $-0.0551(97) $ \\
    1663.3(6)      & 0.0072(25)   & 0            & $0.0072(25)  $ \\
    1672.58(22)      & 0.0186(26)   & 0.019(19)    & $0.000(19)   $ \\
    1679.5(8)      & 0.0045(21)   & 0            & $0.0045(21)  $ \\
    1683.92(16)      & 0.0404(35)   & 0.0606(96)   & $-0.020(10)  $ \\
    1686.8(3)      & 0.0097(24)   & 0.0097(97)   & $0.000(10)   $ \\
    1700.60(14)      & 0.0178(25)   & 0.0451(85)   & $-0.0273(89) $ \\
    1718.88(15)      & 0.0441(39)   & 0.073(11)    & $-0.029(12)  $ \\
    1723.24(19)      & 0.0324(34)   & 0.0376(78)   & $-0.0052(85) $ \\
    1736.01(8)      & 0.0708(48)   & 0.071(36)    & $0.000(36)   $ \\
    1743.72      & 0.0190(30)   & 0.019(12)    & $0.000(12)   $ \\
    1749.1(7)      & 0.0065(25)   & 0            & $0.0065(25)  $ \\
    1756.18(24)      & 0.0189(29)   & 0.0343(77)   & $-0.0154(82) $ \\
    1765.7(4)      & 0.0123(30)   & 0.0152(37)   & $-0.0029(48) $ \\
    1776.8(4)      & 0.0225(32)   & 0.0151(59)   & $0.0074(67)  $ \\
    1803.0(5)      & 0.0130(33)   & 0            & $0.0130(33)  $ \\
    1818.4(5)      & 0.0137(45)   & 0.0136(91)   & $0.000(10)   $ \\
    1823.5(5)      & 0.0160(49)   & 0.016(12)    & $0.000(13)   $ \\
    1842.12(15)      & 0.0340(40)   & 0.034(19)    & $0.000(19)   $ \\
    1849.7(3)      & 0.0057(39)   & 0.0159(78)   & $-0.0102(87) $ \\
    1859.4(7)      & 0.0055(30)   & 0.0393(41)   & $-0.0338(51) $ \\
    1867.3(8)      & 0.0027(22)   & 0            & $0.0027(22)  $ \\
    1872.6(10)      & 0.0044(29)   & 0            & $0.0044(29)  $ \\
    1879.35(8)      & 0.0514(40)   & 0.0671(74)   & $-0.0157(84) $ \\
    1895.67(11)      & 0.0515(39)   & 0.011(11)    & $0.040(12)   $ \\
    1902.6(5)      & 0.0080(25)   & 0            & $0.0080(25)  $ \\
    1941.32(20)      & 0.0281(30)   & 0            & $0.0281(30)  $ \\
    1947.2(4)      & 0.0110(24)   & 0            & $0.0110(24)  $ \\
    1955.40(16)      & 0.0077(39)   & 0.084(26)    & $-0.076(26)  $ \\
    1964.11(20)      & 0.0229(52)   & 0.0250(54)   & $-0.0021(75) $ \\
    1971.50(8)      & 0.0458(45)   & 0.126(12)    & $-0.080(13)  $ \\
    1987.1(5)      & 0.0149(33)   & 0            & $0.0149(33)  $ \\
    1989.9(5)      & 0.0101(24)   & 0.0171(37)   & $-0.0070(44) $ \\
    1996.72(16)      & 0.0296(31)   & 0.0405(67)   & $-0.0109(74) $ \\
    2006.1(4)      & 0.0160(30)   & 0.016(16)    & $0.000(16)   $ \\
    2018.24(14)      & 0.0282(31)   & 0.048(17)    & $-0.020(18)  $ \\
    2024.0(3)      & 0.0193(32)   & 0            & $0.0193(32)  $ \\
    2040.6(4)      & 0.0132(30)   & 0            & $0.0132(30)  $ \\
    2044.80(15)      & 0.0211(38)   & 0.0538(63)   & $-0.0326(73) $ \\
    2048.40(20)      & 0.0313(33)   & 0            & $0.0313(33)  $ \\
    2065.4(3)      & 0.0162(29)   & 0            & $0.0162(29)  $ \\
    2069.1(5)      & 0.0114(29)   & 0.0072(33)   & $0.0042(44)  $ \\
    2078.16(5)      & 0.1329(70)   & 0.1573(87)   & $-0.024(11)  $ \\
    2082.3(5)      & 0.0127(26)   & 0            & $0.0127(26)  $ \\
    2092.6(12)      & 0.0032(23)   & 0            & $0.0032(23)  $ \\
    2103.12(24)      & 0.0152(26)   & 0            & $0.0152(26)  $ \\
    2109.42(22)      & 0.0119(24)   & 0            & $0.0119(24)  $ \\
    2120.61(18)      & 0.0228(36)   & 0.0079(41)   & $0.0149(54)  $ \\
    2125.49(12)      & 0.0451(35)   & 0.045(27)    & $0.000(27)   $ \\
    2129.66(18)      & 0.0296(29)   & 0.014(11)    & $0.015(11)   $ \\
    2144.02(11)      & 0.0606(44)   & 0.0112(34)   & $0.0494(56)  $ \\
    2162.9(3)       & 0.0165(28)   & 0            & $0.0165(28)  $ \\
    2172.26(11)      & 0.0427(37)   & 0.0357(49)   & $0.0070(62)  $ \\
    2174.96(22)      & 0.0140(26)   & 0.0129(45)   & $0.0011(52)  $ \\
    2183.0(4)      & 0.0156(28)   & 0            & $0.0156(28)  $ \\
    2191.87(20)       & 0.0348(34)   & 0            & $0.0348(34)  $ \\
    2198.72(22)      & 0.0291(33)   & 0            & $0.0291(33)  $ \\
    2204.6(4)      & 0            & 0.047(15)    & $-0.047(15)  $ \\
    2230.5(6)      & 0.0084(24)   & 0            & $0.0084(24)  $ \\
    2236.30(13)      & 0.0398(31)   & 0            & $0.0398(31)  $ \\
    2246.6(3)      & 0.0163(28)   & 0            & $0.0163(28)  $ \\
    2257.7(4)      & 0.0110(31)   & 0            & $0.0110(31)  $ \\
    2264.4(5)      & 0.0094(26)   & 0            & $0.0094(26)  $ \\
    2273.9(7)      & 0.0076(26)   & 0            & $0.0076(26)  $ \\
    2280.4(4)      & 0.0113(26)   & 0            & $0.0113(26)  $ \\
    2297.88(9)      & 0.0683(46)   & 0.0739(96)   & $-0.006(11)  $ \\
    2307.6(6)      & 0.0103(26)   & 0            & $0.0103(26)  $ \\
    2311.4(7)      & 0.0062(25)   & 0            & $0.0062(25)  $ \\
    2322.84(15)      & 0.0314(40)   & 0.0327(49)   & $-0.0013(63) $ \\
    2331.3(14)      & 0.0045(26)   & 0            & $0.0045(26)  $ \\
    2340.2(3)      & 0.0093(43)   & 0            & $0.0093(43)  $ \\
    2351.2(4)      & 0.0056(26)   & 0            & $0.0056(26)  $ \\
    2356.01(17)      & 0.0206(33)   & 0.0292(41)   & $-0.0086(53) $ \\
    2361.2(3)      & 0.0141(26)   & 0            & $0.0141(26)  $ \\
    2368.9(9)       & 0.0030(23)   & 0            & $0.0030(23)  $ \\
    2393.23(20)      & 0.0234(35)   & 0.0398(66)   & $-0.0164(75) $ \\
    2396.07(24)      & 0.0249(35)   & 0            & $0.0249(35)  $ \\
    2403.31(10)      & 0.0497(49)   & 0.033(10)    & $0.017(12)   $ \\
    2412.84(22)      & 0.0207(30)   & 0.0043(33)   & $0.0164(45)  $ \\
    2422.60(24)      & 0.0187(35)   & 0            & $0.0187(35)  $ \\
    2437.1(6)       & 0.0076(50)   & 0            & $0.0076(50)  $ \\
    2446.7(7)      & 0.0082(40)   & 0            & $0.0082(40)  $ \\
    2451.5(9)      & 0.0036(30)   & 0            & $0.0036(30)  $ \\
    2459.0(6)      & 0.0140(52)   & 0            & $0.0140(52)  $ \\
    2462.7(4)      & 0.0115(49)   & 0            & $0.0115(49)  $ \\
    2473.3(4)      & 0.0154(35)   & 0            & $0.0154(35)  $ \\
    2483.2(6)      & 0.0107(35)   & 0            & $0.0107(35)  $ \\
    2493.1(3)      & 0.0175(73)   & 0            & $0.0175(73)  $ \\
    2499.48(22)      & 0.0259(38)   & 0            & $0.0259(38)  $ \\
    2521.4(3)      & 0.0213(35)   & 0            & $0.0213(35)  $ \\
    2543.2(3)      & 0.0170(27)   & 0            & $0.0170(27)  $ \\
    2553.72(18)       & 0.0273(31)   & 0            & $0.0273(31)  $ \\
    2562.3(5)      & 0.0143(31)   & 0            & $0.0143(31)  $ \\
    2565.5(5)      & 0.0119(29)   & 0            & $0.0119(29)  $ \\
    2596.11(19)      & 0.0340(40)   & 0            & $0.0340(40)  $ \\
    2599.2(3)      & 0.0216(36)   & 0            & $0.0216(36)  $ \\
    2605.8(3)      & 0.0206(41)   & 0            & $0.0206(41)  $ \\
    2622.1(5)      & 0.0133(43)   & 0            & $0.0133(43)  $ \\
    2629.1(4)      & 0.0133(39)   & 0            & $0.0133(39)  $ \\
    2644.22(22)      & 0.0381(45)   & 0            & $0.0381(45)  $ \\
    2649.3(5)      & 0.0195(42)   & 0            & $0.0195(42)  $ \\
    2814.3(4)      & 0.0192(36)   & 0            & $0.0192(36)  $ \\
    3009.79(20)      & 0            & 0.0193(40)   & $-0.0193(40) $ \\
    5161.005(21)     & 0            & 5.346(62)    & $-5.346(62)  $ \\

\end{longtable}

\begin{minipage}{\linewidth}
\renewcommand{\footnoterule}{}
\footnotetext[1]{$\gamma$-ray intensity adjusted as described in Sect.~\ref{sec:results.2}.\\}
\end{minipage}

\subsection{\label{sec:results.6}Total radiative thermal-neutron-capture cross section for $^{139}$La($n,\gamma$)\protect\\}

The total radiative thermal neutron-capture cross section $\sigma_{0}$ has been investigated for several combinations of PSF/LD models.  This quantity is obtained using Eq.~(\ref{eq:csform}), where $P_{0}$ is the fractional contribution to $\sigma_{0}$ from the calculated intensity of transitions feeding the ground state from all levels in the quasicontinuum above $E_{c} = 285$~keV.  For each model combination, the sum of the conversion-corrected experimentally-measured cross sections (the numerator in Eq.~(\ref{eq:csform})) from levels below $E_{c}$, together with the primary $\gamma$ ray, directly feeding the ground state is $\sum_{i=1}^{n=7} \sigma_{\gamma_{i}0}^{\rm expt}(1 + \alpha_{i0}) = 7.82(50)$~b.  Here, the summation constitutes the following seven $\gamma$ rays of Table~\ref{tab:gammas}: 30.0, 34.6, 43.8, 63.2, 162.6, 272.4, and 5161.0~keV.  

\begin{table}[ht]
  \centering
  \caption{\label{tab:sims} Total radiative thermal neutron-capture cross sections ($\sigma_{0}$), simulated fractions of transitions from the quasicontinuum to the ground state ($P_{0}$) and mean $s$-wave capture-state radiative widths ($\Gamma_{0}$), corresponding to various combinations of $E1$ PSF and LD models.  Bold $\Gamma_{0}$ values are in closest agreement with the adopted value of 50(2)~meV recommended in Ref.~\cite{mughabghab:06}.  Residuals ($R$) between $\Gamma_{0}$ for a PSF/LD combination and the adopted value are presented in the final column.  The acronyms are explained in the text; the PSF combinations A and B are described in Sect.~\ref{sec:results.9}.  Different nuclear realizations give rise to fluctuations in $\Gamma_{0}$ and $P_{0}$ leading to the reported uncertainties.}
  \begin{tabular*}{0.48\textwidth}{@{\extracolsep{\fill}}lcccc@{}}
    \hline\hline
    PSF/LD & $\sigma_{0}$ (b) & $P_{0}$ & $\Gamma_{0}$ (meV) & $|R|$ ($\sigma$)\\
    \hline
    CTF/BA\footnotemark[1]   &  9.81(94) & 0.203(57) & 146(17) & 5.6\\
    CTF/KMF\footnotemark[1]  &  9.52(76) & 0.179(39) & \textbf{53.1(44)} & 0.6\\
    CTF/GLO\footnotemark[1]  & 	9.47(73) & 0.175(36) & \textbf{44.4(28)} & 1.6\\
    BSFG/BA\footnotemark[1]  &  9.47(77) & 0.174(41) & 194(22) & 6.5\\
    BSFG/KMF\footnotemark[1] &  9.27(68) & 0.157(30) & 70.4(64) & 3.0\\
    BSFG/GLO\footnotemark[1] &  9.25(67) & 0.154(28) & \textbf{58.9(45)} & 1.8\\
    CTF/BA\footnotemark[2]   &  9.68(87) & 0.192(51) & 135(17) & 5.0\\
    CTF/KMF\footnotemark[2]  &  9.45(72) & 0.173(35) & \textbf{51.4(47)} & 0.3\\
    CTF/GLO\footnotemark[2]  &  9.41(70) & 0.169(31) & \textbf{43.6(31)} & 1.7\\
    BSFG/BA\footnotemark[2]  &  9.27(81) & 0.157(50) & 244(39) & 5.0\\
    BSFG/KMF\footnotemark[2] &  9.15(69) & 0.146(35) & 91(11) & 3.7\\
    BSFG/GLO\footnotemark[2] &  9.12(68) & 0.143(33) & 76.4(73) & 3.5\\
    CTF/A\footnotemark[1]    &  9.58(80) & 0.184(44) & 46.2(45) & 0.8\\ 
    CTF/B\footnotemark[1]    &  9.64(86) & 0.189(50) & 47.0(45) & 0.6\\
    \hline\hline
    \footnotetext[1]{Assuming the LD parametrization of Ref.~\cite{vonegidy:05}.}
    \footnotetext[2]{Assuming the LD parametrization of Ref.~\cite{vonegidy:09}.}
  \end{tabular*}
\end{table}

The results presented in Table~\ref{tab:sims} indicate that $P_{0}$, and consequently $\sigma_{0}$, show statistical consistency for all permutations of PSF and LD models considered in this analysis.  All results assume the parity dependence of the LD described in Sect.~\ref{sec:theory}, however, parity-independent calculations give $P_{0}$ consistent with the parity-dependent approach.  Similar findings are reported elsewhere \cite{hurst:14,hurst:15,matters:16}.  This observation permits, in essence, a model-independent determination of $\sigma_{0}$.  From the results listed in Table~\ref{tab:sims} we determine a weighted average for $P_{0} = 0.164(39)$, where our uncertainty represents an arithmetic average of the individual uncertainties.  Special combinations involving the PSF models A and B in Table~\ref{tab:sims} were not considered in the averaging process.  Combining this value with the experimental contribution yields an adopted cross section $\sigma_{0} = 9.36(74)$~b.  Of the overall 7.91\% uncertainty in our result, the experimental uncertainty$-$including a statistical uncertainty quadratically-folded with the systematic uncertainty arising from the normalization of the partial $\gamma$-ray cross-section data$-$dominates at the 6.39\% level, while the uncertainty in $P_{0}$ accounts for a 4.67\% contribution.  This result for $\sigma_{0}$ agrees with the value reported earlier in Sect.~\ref{sec:results.4} of 8.58(50)~b corresponding to the total intensity of all transitions experimentally observed to feed the ground state directly.  Our result is consistent with the currently adopted value of 9.04(4)~b \cite{mughabghab:06} and is largely in agreement with the earlier measurements listed in Table~\ref{tab:sigma0}.  It is worth emphasizing that our mean value follows the higher-trending mean-reported values from the most recent activation measurements \cite{arbocco:13,nguyen:14,panikkath:17}.

\begin{figure}[ht]
  \includegraphics[width=0.5\textwidth]{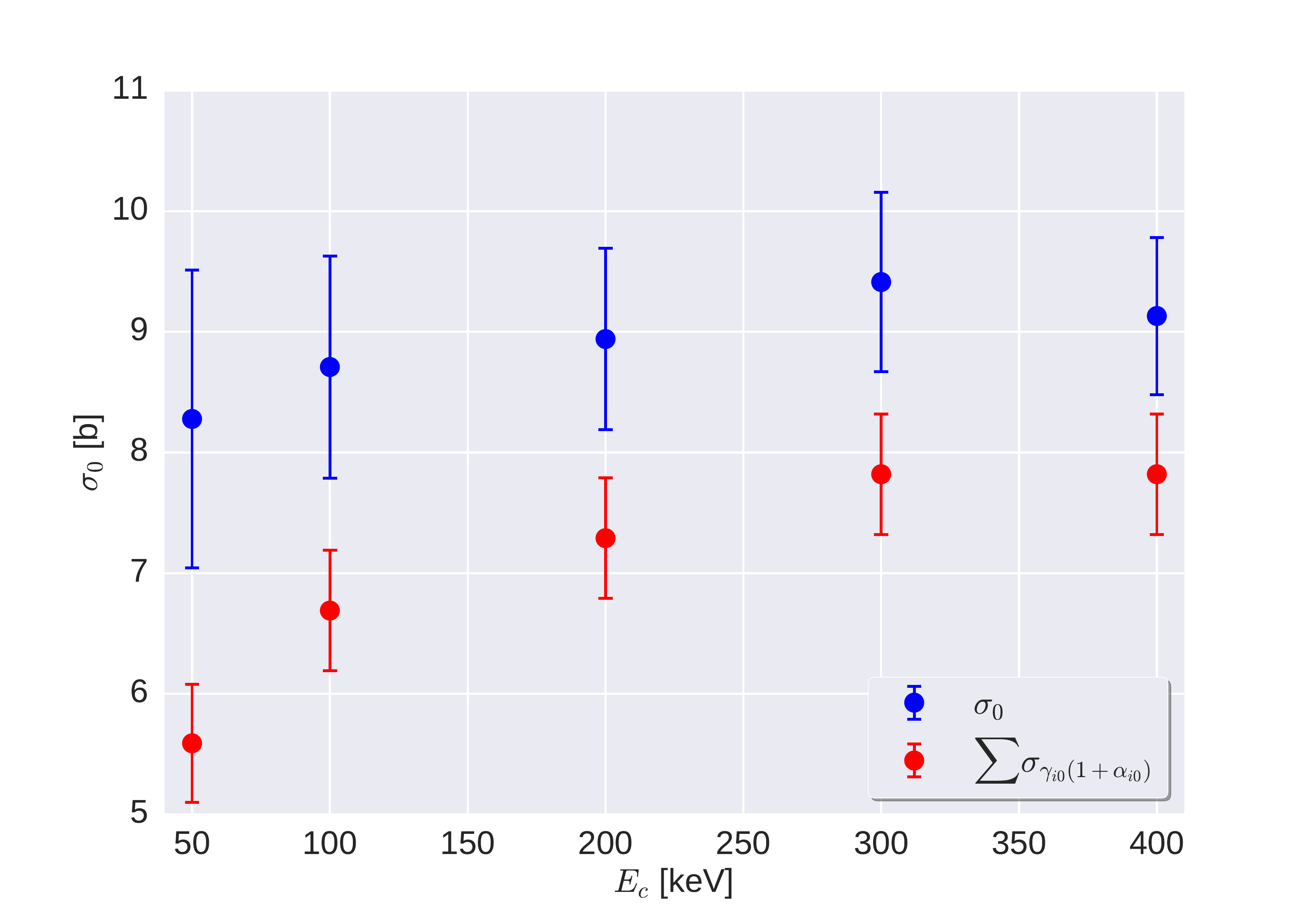}
  \caption{\label{fig:sigma0_VS_Ec} (Color online) Variation in the total radiative thermal neutron capture cross section $\sigma_{0}$ and the sum of the experimental cross sections $\sum_{i} \sigma_{\gamma_{i0}}^{\rm expt}(1+\alpha_{i0})$ feeding the ground state as a function of cutoff energy $E_{c}$.  This plot was obtained with the BSFG/GLO models using the LD parametrization of Ref.~\cite{vonegidy:09} and is representative for PSF/LD combinations adopted in this work.}
\end{figure}

In previous studies of the tungsten \cite{hurst:14,hurst:15} and rhenium \cite{matters:16} isotopes, $\sigma_{0}$ is shown to be a stable quantity with respect to increasing cutoff energy $E_{c}$.  In this work, the $^{139}$La($n,\gamma$) simulations also demonstrate the stability of $\sigma_{0}$ as a function of $E_{c}$, as shown in Fig.~\ref{fig:sigma0_VS_Ec} for the BSFG/GLO model combination for the $E1$ PSF/LD with the LD parametrization of Ref.~\cite{vonegidy:09}.  The features of this plot are representative for any particular PSF/LD combination in Table~\ref{tab:sims}, and shows that both the experimental contribution $\sum_{i} \sigma_{\gamma_{i0}} (1+\alpha_{i0})$ and $\sigma_{0}$ rapidly converge to stable values even before our adopted critical energy ($E_{c} = 285$~keV) is reached.  Even upon extending the cutoff energy beyond this point there is no deviation from stability.

\subsection{\label{sec:results.8}Model discrimination\protect\\}

\begin{figure}[t]
  \begin{center}
    \includegraphics[width=\linewidth]{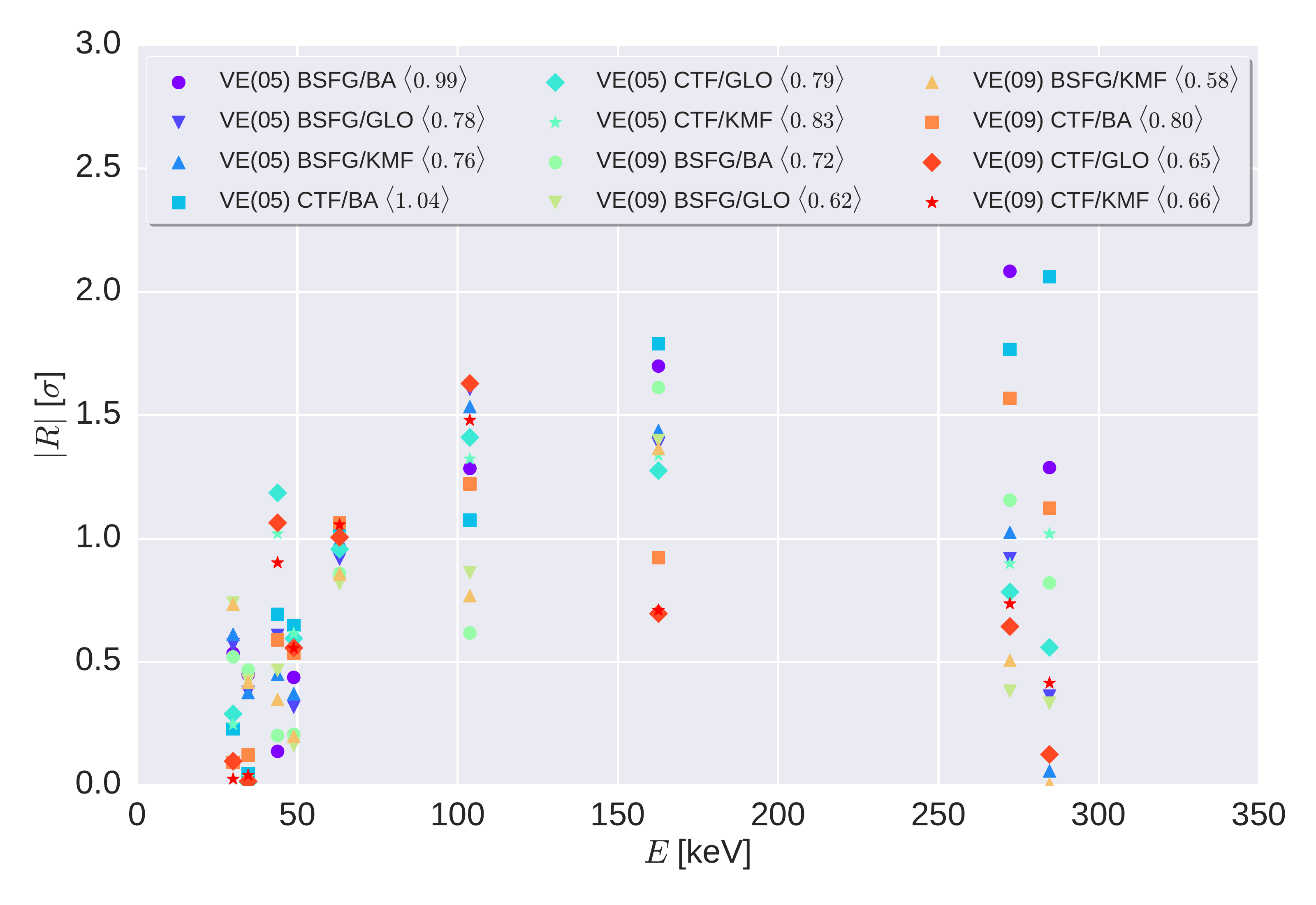}
  \end{center}
  \caption{\label{fig:all_residuals} (Color online) Absolute residuals $|R|$, in units of standard deviations $\sigma$, between simulated-population and experimental-depopulation to all levels below $E_{c}$ for the PSF/LD model combinations adopted in this work (Sect.~\ref{sec:results.6}).  The LD parametrizations for the combinations denoted ``VE(05)'' and ``VE(09)'' are taken from Refs.~\cite{vonegidy:05} and \cite{vonegidy:09}, respectively.  The average residual $\langle R \rangle$ is listed alongside each model combination.}
\end{figure}

All PSF/LD combinations generate a consistent set of $\sigma_{0}$ values and predicted populations to all excited states below $E_{c}$ in good agreement with previous measurements, thus, it is not possible to adopt a preference or rule out any particular combination on this basis alone.  This is evident from the residuals between measured and simulated populations to all levels summarized in Fig.~\ref{fig:all_residuals}.  The only model combinations to show deviations greater than $2\sigma$ are the BA/BSFG and BA/CTF for the 272.3- and 284.7-keV levels, respectively, assuming the LD parametrization of Ref.~\cite{vonegidy:05}.  Overall, it can also be seen from the deduced average residual $\langle R \rangle$ in Fig.~\ref{fig:all_residuals}, that for a given LD model together with its adopted set of parametrizations, combinations invoking the BA PSF produce the largest deviation.

\begin{table}[ht]
  \centering
  \caption{\label{tab:sigma0} Summary of total radiative thermal neutron-capture cross sections for $^{139}$La($n,\gamma$).}
  \begin{tabular*}{0.48\textwidth}{@{\extracolsep{\fill}}llc@{}}
    \hline\hline
    Reference & Method & $\sigma_{0}$ (b)\\
    \hline  
    \textbf{This work}              & \textbf{PGAA} & \textbf{9.36(74)}\\
    Mughabghab \cite{mughabghab:06} & Evaluation    & 9.04(4)\\
    Panikkath \cite{panikkath:17}   & Activation    & 9.28(37)\\
    Panikkath \cite{panikkath:17}   & Activation    & 9.24(25)\\
    Nguyen \cite{nguyen:14}         & Activation    & 9.16(36)\\
    Arbocc{\'o} \cite{arbocco:13}   & Activation    & 9.25(4)\\
    Mannhart \cite{Mannhart}        & Activation    & 8.933(36)\\
    Ryves \cite{Ryves}              & Activation    & 9.03(33)\\
    O'Brien \cite{Brien}            & Chemical separation  & 9.5(5)\\
    Benoist \cite{Benoist}          & Pile oscillator & 8.35(10)\\
    Takiue \cite{Takiue}            & Activation    & 8.63(34)\\
    Gleason \cite{Gleason}          & Activation    & 9.15(25)\\
    Cummins \cite{Cummins}          & Pile oscillator & 9.1(2)\\
    Pomerance \cite{Pomerance}      & Pile oscillator & 8.8(5)\\
    Heft \cite{heft:78}             & Activation    & 9.18(5)\\
    Lyon \cite{Lyon}                & Activation    & 8.10(81)\\
    Harris \cite{Harris}            & Activation    & 9.01(45)\\
    Seren \cite{Seren}              & Activation    & 8.4(17)\\
    \hline\hline
    \end{tabular*}
\end{table}

The agreement between the calculated mean $s$-wave radiative width $\Gamma_{0}$ and the adopted value of $\langle \Gamma_{0} \rangle = 50(2)$~meV \cite{mughabghab:06} may also be used to infer the validity for a particular PSF/LD combination upon inspection of the residuals according to $R = |\langle \Gamma_{0} \rangle - \Gamma_{0}|$.  Table~\ref{tab:sims} clearly indicates a strong model dependency for $\Gamma_{0}$.  The BA model for the PSF, is rather poor at reproducing this quantity regardless of the adopted model for the LD.  The GLO and KMF models fare much better combined with the CTF LD model (indicated in bold font in Table~\ref{tab:sims}), having the closest agreement to the literature value obtained using the KMF/CTF combination for either LD parametrization \cite{vonegidy:05,vonegidy:09}.  Interestingly, the GLO model for the $E1$ PSF also compares well in conjunction with the BSFG LD model assuming the parametrization of Ref.~\cite{vonegidy:05} but not with those of Ref.~\cite{vonegidy:09}.  In addition, it is clear from Fig.~\ref{fig:PSF} that both the KMF and GLO models track the photonuclear \cite{beil:71} and low-energy strength function data \cite{kheswa:15,kheswa:17} rather closely throughout the observed range, while the BA model fails in the low-energy regime.  

The Brink hypothesis states that the shape and size of a resonance is independent of the excitation energy upon which the resonance is built, implying that a model in support of this hypothesis should only exhibit a dependence on $E_{\gamma}$.  Overall, however, our results tend to imply that the commonly adopted BA PSF model, dependent only on $E_{\gamma}$, is less successful at reproducing the observed data.  The GLO and KMF models adopted in this work are better able to reproduce the experimental observables.  Both models of the PSF are based on an additional temperature dependence of the width of the giant dipole resonance, parametrized in terms of the excitation energy of final states populated by $\gamma$-ray transitions as shown in Eqs.~(\ref{eq:temp_width}) and (\ref{eq:temp}), and are consistent with a nonzero limit for the $E1$ PSF as the $\gamma$-ray energy approaches zero.  However, it should be noted that these models are frequently adapted to remove the excitation-energy dependence by utilizing a constant-fit temperature parameter, e.g., Refs.~\cite{voinov:04,siem:02,kheswa:15}.

\subsection{\label{sec:results.9}Modeling the low-energy PSF\protect\\}

Recent investigations carried out in $^{138,139,140}$La by Kheswa \textit{et al}. \cite{kheswa:15,kheswa:17} led to the development of a compound function to describe the overall photon strength in these isotopes.  We have adopted this PSF and its associated parametrizations as described in Sect.~\ref{sec:theory.3} to assess predictions with this model and perform an independent analysis of our ($n,\gamma$) data; these results are also listed in Table~\ref{tab:sims}.  In the first test we modeled this PSF assuming all resonance components correspond to pure $E1$ strength (``Model A'') and in a second test we modeled the lowest-energy resonance as an $M1$ contribution (``Model B'').  For each case, we used the CTF LD model together with the parametrization of Ref.~\cite{vonegidy:05}, and assumed an additional single-particle contribution to the $M1$ PSF of $5 \times 10^{-9}$~MeV$^{-3}$.  Both simulations yield $P_{0}$ (and, therefore, $\sigma_{0}$) values consistent with others listed in Table~\ref{tab:sims}.  Because the adopted literature value for $\langle \Gamma_{0} \rangle$ \cite{mughabghab:06} is used to normalize the data reported in Refs.~\cite{kheswa:15,kheswa:17}, values for $\Gamma_{0}$ consistent with this expectation value are regenerated in these calculations.  Thus, it is not possible to pin down the electric- or magnetic-dipole characterization of the PSF at low energies from this analysis since both give comparable results.  However, it is interesting to note that observables produced using traditional phenomenological models of the PSF adopted in this work, based on a parametrization of the giant dipole resonance at high energy to describe the $E1$ strength, are fully consistent with the empirical function \cite{kheswa:15} deduced using the totality of the available low-, mid- and high-energy data.

In the context of systematics approaching the $N=82$ shell closure, we find the PSF for $^{140}$La ($N=83$) is best described using models exhibiting a similar response to that observed in the reportedly near-spherical $^{148}$Sm ($N=86$) rather than modestly-deformed $^{149}$Sm ($N=87$) \cite{siem:02}.  Although the experimental data for $^{138}$La (Fig.~\ref{fig:PSF}) appears to reveal a mild enhancement at the very lowest energies, a variance analysis of the strength function data below 5.2~MeV for all $^{138,139,140}$La isotopes reveals consistency at the 95\% confidence level (CL).  Furthermore, the empirical distribution functions for these data sets agree with the continuous cumulative distribution for the GLO function at ${\rm CL \geq 85\%}$ over the same energy interval, $E_{\gamma} < 5.2$~MeV.  Neighboring nucleus $^{144}$Nd ($N=84$) \cite{voinov:15} also displays similar characteristics at low energy to the data for $^{138-140}$La and $^{148}$Sm.  These observations may lend support to the shape-transitional claim proposed in the samarium isotopes \cite{siem:02} because $^{138-140}$La ($|\beta_{2}| \leq 0.045$ \cite{moller:95}) and $^{144}$Nd ($\beta_{2} = 0$ \cite{moller:95}) are well characterized as near-spherical systems.

\section{\label{sec:summary}Summary\protect\\}

A set of partial $\gamma$-ray cross sections has been measured for the $^{139}$La($n,\gamma$) reaction using thermal neutrons.  These cross sections are combined with {\small DICEBOX} statistical-model calculations to yield the total radiative thermal neutron-capture cross section, $\sigma_{0} = 9.36(74)$~b.  This cross section is consistent with the sum of experimentally-observed transitions feeding the ground state directly: $\sum_{i} \sigma_{\gamma_{i0}}(1+\alpha_{i0}) = 8.58(50)$~b.  Because it is expected that a fraction of ground-state transitions remain unobserved, it is not surprising that the experimental sum is lower.  Our result for $\sigma_{0}$ agrees with the recent higher-trending activation measurements of 9.28(37)~b \cite{panikkath:17}, 9.24(25)~b \cite{panikkath:17}, 9.25(4)~b \cite{arbocco:13} and 9.16(36)~b \cite{nguyen:14}, as well as the earlier adopted value of 9.04(4)~b \cite{mughabghab:06} based on other measurements listed in Table~\ref{tab:sigma0}.  The experimental cross sections reported in this work provide new information for the Evaluated Gamma-ray Activation File (EGAF) \cite{firestone:06} and the Experimental Nuclear Reaction Data (EXFOR) \cite{otuka:14} databases that will help guide and improve future evaluations for the Evaluated Nuclear Data File (ENDF) \cite{brown:18}.  The new $\gamma$-ray data will also be useful for the International Reactor Dosimetry and Fusion File (IRDFF) \cite{zsolnay:12} standardized cross-section library of neutron dosimetry reactions used in a range of neutron-metrology applications.

In this study, we are able to confirm spin-parity assignments \cite{nica:07} for eight levels levels below our established value of $E_{c} = 285$~keV, and suggest a revised $J^{\pi} = 5^{-}$ assignment for the 103.8-keV level.  Also, because the predicted populations agree well with the measured depopulation data up to $E_{c}$, this is further evidence for decay-scheme accuracy and completeness.  Furthermore, our value for $E_{c}$ represents a significant increase over the number of levels in the RIPL file up to which the spins and parities are uniquely assigned at 63.2~keV (i.e. the first six levels) \cite{capote:09}.  Above $E_{c}$, we deduce $J^{\pi} = 6^{-}$ for the 322.0-keV level corresponding to an essentially pure $\pi(2d_{5/2})\otimes\nu(2f_{7/2})$ configuration, consistent with the dominance of $l=3$ transfers \cite{kern:67} and the expected multiplet of states in the absence of particle-phonon coupling below 600~keV.  The decay-scheme analysis also provides evidence for an additional eight unique $J^{\pi}$ assignments above this excitation energy, while confirming 23 adopted assignments (or ranges).

From the spectroscopic analysis of the prompt primary $\gamma$-ray data, we are able to confirm previous experimental observations \cite{islam:90} and provide a new independent measurement of the neutron separation energy in $^{140}$La, $S_{n} = 5161.005(21)$~keV.  This result is consistent with, yet has a smaller uncertainty than, the present adopted value \cite{wang:17}, and provides useful input for future atomic mass evaluations.  Furthermore, we are able to constrain spin windows for 123 levels populated by primary $\gamma$ rays.

Finally, the statistical-model analysis shows that PSF models with $E1$ strength corresponding to a nonzero limit as the $\gamma$-ray energy approaches zero are better able to reproduce the recommended width of the neutron resonances just above $S_{n}$ \cite{mughabghab:06}, as well as the experimental photonuclear \cite{beil:71} and low-energy strength function data from Oslo-type measurements \cite{kheswa:15,kheswa:17}.  However, all PSF/LD model combinations generate consistent predictions for $P_{0}$ in $^{140}$La.

\section*{Acknowledgments\protect\\}

This material is based upon work supported by the Department of Energy National Nuclear Security Administration under Award Numbers DE-NA0003180 and DE-NA0000979.  This work is also supported by the Lawrence Berkeley National Laboratory under Contract No. DE-AC02-05CH11231 for the US Nuclear Data Program, and the Lawrence Livermore National Laboratory under Contract no. DE-AC52-07NA27344.  Additional support was received through the Undergraduate Research Apprentice Program and the Nuclear Science and Security Consortium.  We thank the Budapest Neutron Centre access program for funding the PGAA beamtime.  L.~S. acknowledges the financial support of the J{\'a}nos Bolyai Research Fellowship of the Hungarian Academy of Sciences, as well as the Project No. 124068 of the National Research, Development and Innovation Fund of Hungary, financed under the K17 funding scheme.  A.~H. thanks Dr.~M.~Guttormsen and Dr.~A.~C.~Larsen for helpful discussions and reviewing the manuscript.

\bibliography{140La}

\end{document}